\documentclass[a4paper,fleqn]{cas-sc}

\usepackage[numbers]{natbib}
\usepackage{amsmath,amssymb,amsfonts}
\usepackage{graphicx}
\usepackage{textcomp}

\usepackage{algorithm}  
\usepackage{algorithmicx}  
\usepackage{algpseudocode}  
\usepackage{dirtytalk}
\usepackage{subcaption}
\usepackage{accents}

\usepackage{soul}

\usepackage{color,soul}

\usepackage{hyperref}
\usepackage{booktabs} 
\usepackage{placeins}
\usepackage{multicol}
\usepackage[para,online,flushleft]{threeparttable}

\usepackage{tikz}

\usepackage{algorithm}
\usepackage{algpseudocode}
\algnewcommand\algorithmicforeach{\textbf{for each}}
\algdef{S}[FOR]{ForEach}[1]{\algorithmicforeach\ #1\ \algorithmicdo}

\begin{document}
\let\WriteBookmarks\relax
\def\floatpagepagefraction{1}
\def\textpagefraction{.001}

\shorttitle{TMS-Net}    

\shortauthors{F. Uslu and A. Bharath}  

\title [mode = title]{TMS-Net: A Segmentation Network Coupled With A Run-time Quality Control Method For Robust Cardiac Image Segmentation}


\tnotemark[1]

\author[1]{Fatmat\"{u}lzehra Uslu}[orcid=0000-0001-7153-7583]

\cormark[1]




\ead{fatmatulzehra.uslu@btu.edu.tr}

\credit{Conceptualization of this study, methodology, doing analysis, writing the paper}

\affiliation[1]{organization={Bursa Technical University},
            addressline={Electrical and Electronics Engineering Department}, 
            city={Bursa},
            postcode={16310}, 
            country={T\"{u}rkiye}}

\author[2]{Anil A. Bharath}[orcid=0000-0001-8808-2714]




\ead{a.bharath@imperial.ac.uk}

\credit{Editing the paper}

\affiliation[2]{organization={Imperial College London},
            addressline={Bioengineering Department}, 
            city={London},
            postcode={SW7 2AZ}, 
            country={UK}}
            
\cortext[1]{Corresponding author}



\begin{abstract}
Recently, deep networks have shown impressive performance for the segmentation of cardiac Magnetic Resonance Imaging (MRI) images. However, their achievement is proving slow to transition to widespread use in medical clinics because of robustness issues leading to low trust of clinicians to their results. Predicting run-time quality of segmentation masks can be useful to warn clinicians against poor results. Despite its importance, there are few studies on this problem. To address this gap, we propose a quality control method based on the agreement across decoders of a multi-view network, TMS-Net, measured by the cosine similarity. The network takes three view inputs resliced from the same $3D$ image along different axes. Different from previous multi-view networks, TMS-Net has a single encoder and three decoders, leading to better noise robustness, segmentation performance and run-time quality estimation in our experiments on the segmentation of the left atrium on STACOM $2013$ and STACOM $2018$ challenge datasets. We also present a way to generate poor segmentation masks by using noisy images generated with engineered noise and Rician noise to simulate undertraining, high anisotropy and poor imaging settings problems. Our run-time quality estimation method show a good classification of poor and good quality segmentation masks with an AUC reaching to $0.97$ on STACOM $2018$. We believe that TMS-Net and our run-time quality estimation method has a high potential to increase the thrust of clinicians to automatic image analysis tools. 
\end{abstract}


\maketitle

\section{Introduction}

Although deep networks provide promising solutions to reduce workload of clinicians for quantitative studies \cite{cao2020application,zhou2017quantitative} or further region-based processing \cite{kerfoot2018left,uslu2021net,yang2018multiview}, the integration of these models into clinical workflows is not at a desired level. One of the barriers for the delayed integration is the lack of trust of clinicians in the outputs of these models \cite{lu2021clinical}. All artificial intelligence (AI) systems can produce incorrect results for either segmentation or classification. The purpose of this work is to suggest an approach which performs a form of self-consistency checking which helps provide a confidence measure to network outputs for medical image analysis. The approach is developed, in this paper, for image segmentation.

Errors in the output of a trained network can occur due to issues related to model training. When a deep model is trained with a small dataset that does not reflect the prior distribution of classes, or diverging from the statistical properties of data seen during training, the model may not generalise well to unseen images. This issue usually exacerbates with the deployment set to be, somehow, different from training set, such as variations in vendors of the same imaging modality or collecting data from cohorts with different properties. Also, malfunctioning imaging devices can contribute to this problem \cite{krittanawong2019deep,lu2021clinical}. Moreover, some $3D$ medical images suffer from high anisotropy due to large slice thickness, which may limit model performance due to limited information along the out-of-plane axis \cite{jia20193d,zeng2017deepem3d}. The most significant weakness of deep networks can be reporting high confidence on their outputs, which is preserved even when generating incorrect results. Without improved automatic measures of confidence in network outputs, their acceptance to clinical routine may be delayed.

In the literature, there are some approaches used for improving trustworthiness of deep models to some extent, such as using attribution maps to understand which locations/features are used in decision making, network dissection to explore which neurons are responsible for particular decisions, confidence calibration to transform saturated network outputs to class probabilities, and uncertainity estimation with Monte Carlo (MC) dropouts or ensemble models to measure confidence of networks on their outputs  \cite{salahuddin2022transparency,mehrtash2020confidence,gawlikowski2021survey}. While these methods give some clues for interpretation of how a model reaches its particular decision or its confidence on its outputs, they may not always provide a direct way for automating the elimination of poor segmentation results, by still requiring an expert to interpret them. On the other hand, recent literature has presented classifier or a regression based models for detecting failure cases \cite{hann2021deep,frounchi2011automating,roy2018inherent}. Regression based methods predict run-time performance by training a regression model with features extracted from various qualities of segmentation masks, whose outputs are, later, thresholded to obtain binary results of "accept" or "reject". However, these methods, themselves, require large datasets for training to avoid learning a biased decision. To deal with it, some work generated a range of quality segmentation masks by various degraded segmentation models \cite{hann2021deep,roy2018inherent,robinson2019automated}. However, this also has the disadvantage of long training time due to training multiple copies of the same model to generate various qualities of segmentation masks.

Some previous studies used model uncertainty obtained with MC dropouts or ensemble models to predict poor segmentation masks \cite{hann2021ensemble,wang2019aleatoric,gawlikowski2021survey}. The former method uses dropouts -- originally proposed as a remedy to model overfitting by randomly disabling some connections in a network during training -- to approximate Bayesian inference during test time \cite{gal2016dropout}. On the other hand, the latter method trains multiple models for the same task, where multiple copies of the same network with different initialisation points or different architectures can be used \cite{hann2021deep}. Despite being computationally more expensive, ensemble models have been found to exhibit better performance at confidence calibration or run-time quality estimation \cite{hann2021ensemble,jungo2019assessing,ovadia2019can}. Multi-view networks, analysing a $3D$ image with three $2D$ networks, can be viewed as a ensemble model, where three copies of the same architecture are independently trained to approximate segmentation maps of three views of the same organ, sampled from the same image data \cite{mortazi2017cardiacnet}. As far as we are aware, no previous work has explored multi-view networks for run-time quality estimation.

In this study, we propose a trustworthy multi-view segmentation network for medical image segmentation, called TMS-Net, standing for \textbf{T}rustworthy \textbf{M}ulti-view \textbf{S}egmentation \textbf{Net}work. In addition to generating accurate segmentation results, agreement across TMS-Net outputs can be used for run-time quality estimation. Therefore, a human operator can be alerted to review its segmentation results when potentially low quality segmentation is produced. Multiview networks, taking the same $3D$ volume sliced along different axes, are usually used instead of $3D$ networks, due to the lower memory and computational requirements during training \cite{mortazi2017cardiacnet}. In this study, we explore the possibility of multi-view networks to be used for confidence estimation, similar to deep ensemble models. Based on previous work, reporting sharing low-level parameters improves performance of ensemble networks \cite{lee2015m,yu2020ensemble}, TMS-Net uses a single encoder connected to three decoders. This specific design also provides a common base for measuring similarities between decoder outputs by making decoders dependent on the same coding function. On the other hand, we independently train decoders with a particular view of a $3D$ MRI image to allow each decoder to decode different representation of the same volume image, similar to previous work using specialised loss functions to diversify network outputs \cite{lee2015m,yu2020ensemble}.

In this work, we also present a run-time quality score based on cosine similarities to quantify agreement between TMS-Net outputs, in contrast to previous work training a regression or classification model for the same aim \cite{frounchi2011automating,roy2018inherent,hann2021ensemble}. Previous work trained degraded versions of the same model to generate a range of quality segmentation masks for training or performance evaluation of their methods \cite{hann2021ensemble,robinson2019automated,valindria2017reverse}, by addressing undertraining problem of supervised methods. In contrast, in this work, we add various magnitudes of engineered or synthetic Rician noise to input images for the same aim, similar to the work of Gheorghi\c{t}\v{a} \textit{et al.} \cite{gheorghitua2022improving} generating synthetic data to simulate patient characteristics which are not represented in training set. With the corruption of volume images with engineered noise at three views of the sagittal, axial or coronal, or a single view, or by reducing the visibility of structures in images with Rician noise, we are able to respectively assess model performance in the presence of insufficiently small dataset, high anisotropy or poor imaging settings.  We evaluate the segmentation and run-time quality estimation performance of our methods on the segmentation of left atrium (LA) in cardiac MRI images of STACOM $2013$ and $2018$ datasets.

Our contributions:
\begin{enumerate}
\item We present a multi-view network, TMS-Net, showing more robustness to several problems, such as insufficient size of training data, high anisotropy or poor imaging settings, compared to other segmentation networks.
\item We simulate poor segmentation by corrupting the original images with engineered and Rician noise, instead of training multiple copies of the same network. 
\item We propose an unsupervised run-time quality estimation method. 
\end{enumerate}

The remainder of this paper is organized as follows. Section $2$ gives some background information about the deep models proposed for the segmentation of $3D$ medical images, and run-time quality estimation methods. Section $3$ describes the design of TMS-Net and our run-time quality estimation method. Section $4$ gives the details of our experiments and datasets used in the paper. Section $5$ presents the results for our run-time quality estimation method and compares the segmentation results of TMS-Net with previous methods on the original and corrupted datasets. Finally, Section $6$ provides a discussion about the results of our experiments.

\section{Background}
Although $3D$ MRI is very common, much of the literature on segmentation is based around slice-wise networks. This is partly because training intrinsically $3D$ segmentation networks is very computationally expensive in terms of labelling,  data and GPU memory. However, because it is easier to train $2D$ networks slice by slice, this gives us the opportunity to exploit the different views for segmentation to derive an efficient new confidence score on segmentation results. Among $2D$ segmentation networks, U-Net \cite{ronneberger2015u} and SegNet \cite{badrinarayanan2017segnet} are commonly used so we compare the performance of our network with them. U-Net is one of the most commonly used segmentation networks for medical image analysis, which is an encoder-decoder architecture with skip connections \cite{ronneberger2015u}. Its encoder consists of double convolutional layers followed by downsamplig layers by a factor of $2$. On the other hand, its decoder is symmetrical to its encoder, where upsampling layers replace donwsamplig layers. To increase details in decoder features, skip connections carry encoder features to the decoder path. In contrast to U-Net \cite{ronneberger2015u}, SegNet \cite{badrinarayanan2017segnet} keeps max pooling indices during downsampling of encoder features and reuse them later in decoder features when upsampling them. There are various derivatives of the U-Net\cite{ronneberger2015u} proposed to improve its performance on several segmentation tasks  such as adding attention modules to improve the network's ability to focus on a specific organ \cite{oktay2018attention} or increasing feature variety/training stability by replacing double convolutional layers with inception networks, dense layers or residual layers \cite{kerfoot2018left, zhou2018unet, zhang2020dense}. More details can be found in the survey of Siddique \textit{et al.} \cite{siddique2021u}.

For the completeness of the literature on the segmentation of $3D$ images, we will briefly review $3D$ networks. More details can be found on the survey of Singh \textit{et al.} \cite{singh20203d}. The main distinction between $2D$ and $3D$ networks is the use of $3D$ convolutional layers in the latter one, which facilitates to effectively use inter-slice dependencies in segmentation. Examples to the most known $3D$ networks are $3D$ U-Net \cite{cciccek20163d}, V-Net \cite{milletari2016v}, DeepMedic \cite{kamnitsas2016deepmedic}, and nnU-Net \cite{isensee2021nnu}. Although $3D$ networks are a better choice for volumetric image analysis in many problems, they have some disadvantages. Firstly, they require a larger size of training set because of their large parameter counts due to the use of $3D$ filters. As a result of being larger models, $3D$ networks can lead to memory problems given the high resolution of medical images. Previous studies dealt with that by down-sampling input images, which can be result in missing details in input images, or training networks with sub-volumes \cite{singh20203d}. Moreover, highly anisotropic images may not be a good candidate for $3D$ networks due to limited representation of anatomical structures along the $z$ axis.

A trustworthy segmentation method is expected to have high confidence in its output when it is accurate, which generally happens when its input image is similar to examples in the training dataset, and vice verse. However, current networks can generate high response even when they are wrong. Therefore, it is important to check their outputs before using them in medical clinics or automatic image analysis pipelines. Recent run-time quality estimation methods were mostly based on measuring consistency between a segmentation mask, whose quality is to be estimated, and a pseudo ground truth, which can be segmentation samples generated by MC dropouts or ensemble models \cite{frounchi2011automating,roy2018inherent,hann2021ensemble}. Frounchi \textit{et al.} \cite{frounchi2011automating} used a \textit{test oracle} to pick poor segmentation results during run-time, where high quality segmentation masks were associated with high similarities to those whose accuracy was validated medical experts in the beginning or with well performing segmentation method in later stages. They trained a classifier with volume difference, geometrical and overlap based metrics, obtained from segmentation masks, to identify poor segmentation masks. Valindria \textit{et al.} introduced the concept of \textit{reverse classification accuracy, (RCA)} for run-time quality estimation, and applied it to multi-organ segmentation \cite{valindria2017reverse}; the method was later validated on cardiac Magnetic Resonance Imaging (MRI) images in the UK Biobank dataset by Robinson \textit{et al.} \cite{robinson2019automated}. In this method, a segmentation model, called \textit{reverse classifier}, is trained with an input image and its predicted segmentation mask, let's say $M$, whose accuracy is desired to be measured. Then, the trained segmentation model is assessed on a reference set, which contains images and their ground truth segmentation masks, which is expected to be sufficiently diverse to well represent possible cases. The maximum segmentation performance on the reference set suggests the run-time performance of $M$. Despite of its high performance, this method was reported to be sensitive to type of segmentation model and has long execution time \cite{valindria2017reverse}. Roy \textit{et al.} \cite{roy2018inherent} trained a regression model to predict ground truth Dice scores from mean Dice score calculated over MC samples, where they found a strong correlation between these parameters. However, the performance of their method was reported to moderately vary on different brain MRI datasets, by showing sensitivity to dataset characteristics. In addition to MC dopouts, Hann \textit{et al.} \cite{hann2021ensemble} produced samples with ensemble models for run-time performance estimation on cardiac MRI image segmentation. They reported a high classification rate of $99 \%$. As long as we are aware, there is no previous work on run-time quality estimation of left atrium segmentation in MRI images.

\section{Method}

\begin{figure}[!ht]
\centering
\includegraphics[width=1 \textwidth]{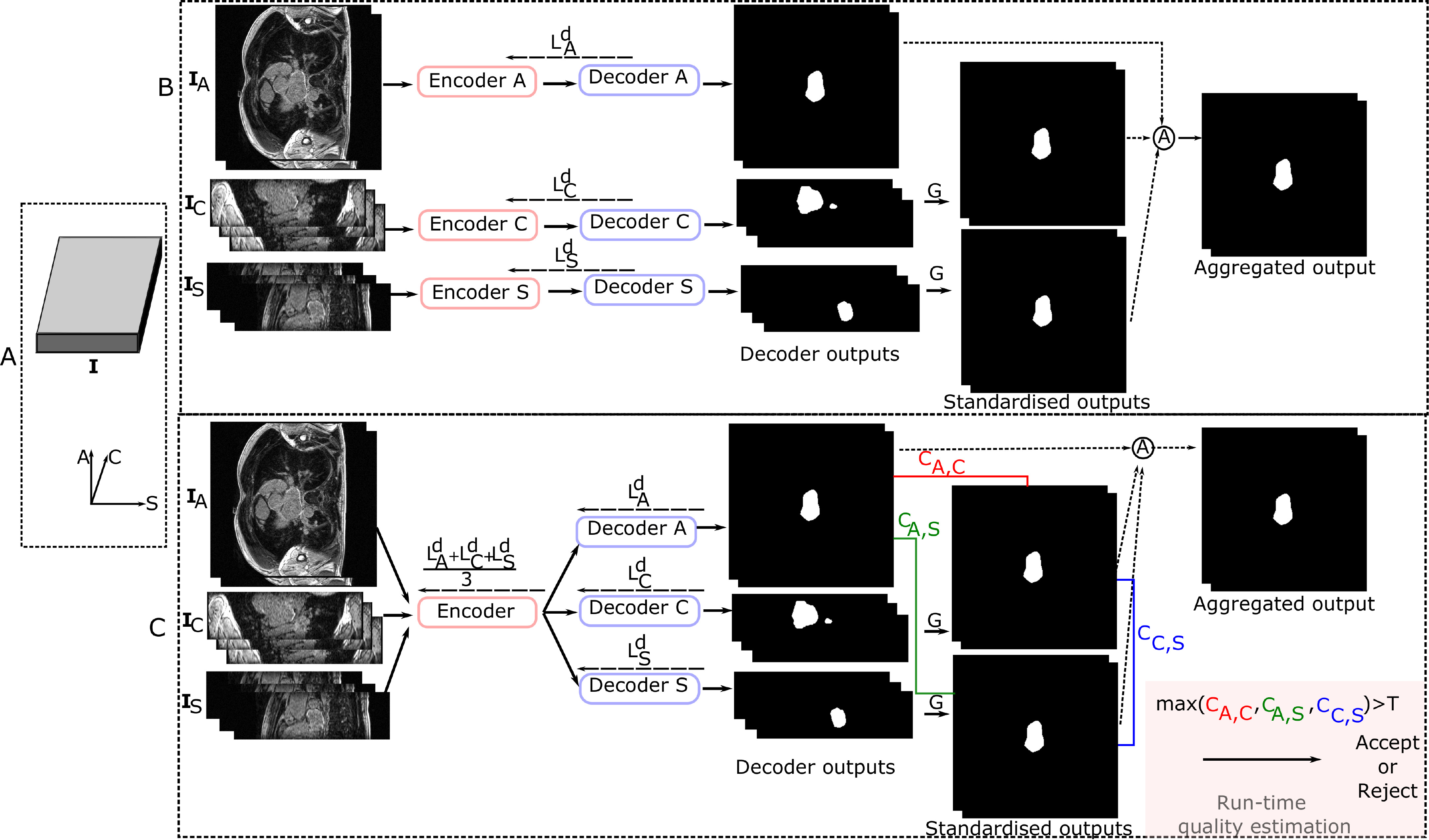}
\caption{(A) Resampled views from a $3D$ volume image, where light gray/dark gray/black surfaces represent the axial (A)/coronal (C)/sagittal (S) view images. The coordinate system in the bottom shows sampling directions for these images. (B) A general design of multi-view networks \cite{yang2018multiview,mortazi2017cardiacnet}, (C) A general schematic of TMS-Net coupled with our run-time estimation method.  TMS-Net is a single network with a shared encoder and three decoders in contrast to the design shown in (B) with three independent subnetworks. Our run-time estimation method generates consistency scores $C_{\cdot, \cdot}$ between the standardised outputs of a decoder pair to accept or reject the aggregated segmentation mask. In (B) and (C), function $G$ means reslicing a segmentation mask volume to obtain a standardised output, which has the axial view in the figure. \textcircled{\raisebox{-0.9pt}{$A$}} denotes an aggregation function to combine the output of each sub-network to synthesise the aggregated output. Solid and dashed arrows respectively show forward and backward propagation. The encoders of subnetworks in (B) are independently trained with losses associated with their decoder outputs; however, TMS-Net uses the mean loss calculated over its three decoder outputs to train its encoder.  See text for more details. (Best viewed in color.)} \label{fig:Pipeline}
\end{figure}

Although multi-view networks \cite{mortazi2017cardiacnet} are designed to be as cost-effective substitutes of $3D$ networks in medical image analysis (see Figure \ref{fig:Pipeline}), they mostly ignore dependency between three views of the same $3D$ image by training independent networks for the views. In other words, images resampled along the axial, coronal and sagittal views, $I_{A}$, $I_{C}$ and $I_{S}$, are treated as if they correspond to different source images, by ignoring shared image characteristics across them, such as noise type in a volume image. Training a shared encoder with images of the three views can encode common and specific information to each view. This constraint also provides the basis for our run-time performance estimation technique, explained as follows.

\subsection{TMS-Net} \label{sec:TMS-Net}

As its general structure is demonstrated in Figure \ref{fig:Pipeline}, TMS-Net \footnote{Our code is available at \url{https://github.com/fzehrauslu/TMS-Net}} is a multi-view network with a shared encoder and three decoders. Our network adopts the design of image restoration networks using the wavelet analysis techniques, to promote noise robustness \cite{liu2019multi}. Its encoder contains wavelet analysis modules to extract features within various frequencies. The details of TMS-Net architecture will be explained as follows.

\begin{figure*}[!ht]
\centering
\includegraphics[width=0.99\textwidth]{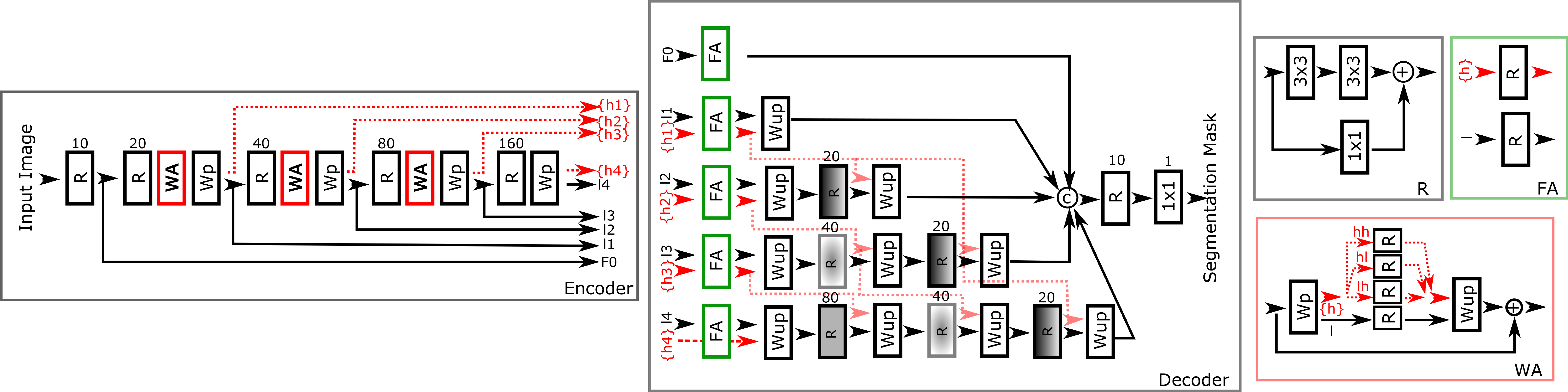}
\caption{Details of TMS-Net. Refer to Figure \ref{fig:Pipeline} for its overall structure. \textit{R}:residual module, \textit{WA}:wavelet analysis module, where filter counts of residual modules are equal to the number of input features, \textit{Wp}: wavelet pooling layer. Residual layers with matching gray patterns in the decoder share parameters. (Best viewed in color.) } \label{fig:TMS-Net}
\end{figure*}

\subsubsection{Encoder }
TMS-Net encoder contains five residual \cite{he2016deep} and three wavelet analysis modules, and four wavelet pooling layers (see Figure \ref{fig:TMS-Net}).

\paragraph{Wavelet Analysis Module }

This module contains a pair of pooling and unpooling layers and four parallel residual modules between them. The design of this module is inspired by classical wavelet analysis methods \cite{mohan2014survey}, where an input image is firstly decomposed into low and high frequency components (see Section \ref{sec:WaveletAnalysis} in Appendix for more details). In the classical approach, after thresholding high frequency components, assumed to contain noise, the low and high frequency components are sent through the wavelet reconstruction to produce a denoised image. Similar to the classical approach, we firstly decompose input features with the wavelet pooling layer into low and high frequency features, $f_{ll}$ and, $f_{lh}$, $f_{hl}$ and $f_{hh}$. Then, these features are independently processed with independent residual modules, prior to sent to the wavelet unpooling layer. Our module also contains a residual connection to facilitate stable training, which is commonly used in the design of recent deep denoising networks \cite{jifara2019medical,li2020mri,tripathi2020cnn}.

\paragraph{Wavelet Pooling Layers}

We use wavelet pooling layers to downsample features. These layers perform wavelet decomposition, which naturally downsamples inputs by factor of $2$ and produce low and high frequency components/features of inputs. Only low frequency ones are passed down along the encoder path to generate multi-scale features while both feature types are sent to decoder path through skip connections. See Figure \ref{fig:TMS-Net} for details, where skip connections carrying low frequency features are shown in black and those passing high frequency features are displayed in red.

\subsubsection{Decoder}
TMS-Net has three decoders, each corresponding to a particular view (see Figure \ref{fig:Pipeline}(b)). Each decoder contains five parallel branches, which are connected to a residual module in the end of the decoder (see Figure \ref{fig:TMS-Net}). Each branch takes a different resolution of encoder features through skip connections and gradually increases it to the resolution of input images, with wavelet unpooling. Depending on the resolution of encoder features, the lengths of branches vary. Each branch starts with a feature adaptation module and usually continues with alternating layers of wavelet unpooling and residual layers.

Because wavelet unpooling layers are non-trainable, residual layers and feature adaptation modules make the decoder branch trainable. Residual layers with the same feature resolution share parameters to avoid learning any redundant features. These residual layers are illustrated with matching gray patterns in Figure \ref{fig:TMS-Net}. In contrast to residual layers receiving only low frequency features, feature adaptation modules take both high and low frequency features as input and independently analyse them with two parallel residual layers. Therefore, they can be viewed as adapting encoder features of various resolutions to related branches, in the beginning of a decoder.

For simplicity, one can use an averaging operator, as the aggregation function $\textcircled{{\small{A}}}$ (see Figure \ref{fig:Pipeline}), to combine the outputs of decoders. Because the decoders generate output volumes resliced along specific views, one needs to bring them to the same view. For this reason, we select a standard view and reslice decoder outputs to match their representation with that of the standard view. Eq. (\ref{eq:aggregation}) calculates aggregated decoder outputs:
\begin{equation}\label{eq:aggregation}
g''(V_{A},V_{C},V_{S})=\frac{1}{3}\big(G(V_{A})+G(V_{C})+G(V_{S})\big)
\end{equation}
where $G(\cdot)$ is composite function that, firstly, transforms a volume generated at a particular view to a standard view, then apply a smoothing function.

\subsubsection{Training Steps of TMS-Net } 
Because of the constraint in the new design, the sub-networks can not be trained independently anymore; therefore, we present a training strategy tailored for our network (See Algorithm \ref{alg:cap}). Each training epoch has four parts, where we start with training the encoder and continue with consecutive training of decoder $A$, $C$ and $S$. When optimising a specific part with a loss function in Eq. (\ref{eq:LossFunc1} -\ref{eq:LossFunc2}), we freeze the parameters of the rest of the network. The two loss functions that we used for TMS-Net are 
\begin{align} 
& L^{e} =\frac{1}{3}(L_{A}^{d}+L_{C}^{d}+L_{S}^{d}) \label{eq:LossFunc1}\\
& L_{V}^{d}= \frac{1}{N_{V}} \sum_{n}^{N_{V}}\mathit{l}_{c}(S_{V}^{n},\tilde{S_{V}^{n}}) \quad V \in \{A,C,S\}\label{eq:LossFunc2}  
\end{align}
where $L^{e}$ shows the mean loss computed for the encoder, which takes input from the three views, similar to multi-task training \cite{mao2020multitask}.  $L_{V}^{d}$ is the mean loss calculated for a particular decoder matched with an orthogonal view $V$, and $N_{V}$ represents the number of training samples for the same view. $\mathit{l}_{c}$ denotes the cross entropy loss. $S_{V}$ corresponds to a $2D$ ground truth segmentation mask and $\tilde{S_{V}}$ shows its estimation.

\begin{algorithm}
\caption{Training Procedure of TMS-Net.}\label{alg:cap}
\begin{algorithmic}[1]
\ForEach {epoch}
\State Freeze only the parameters of decoders $A$, $S$ and $C$; then, train the encoder with the loss $L^{e}$, in Eq. (\ref{eq:LossFunc1}) \footnotemark \Comment{At this step, input images $I_{A}$, $I_{C}$ and $I_{S}$ are randomly selected and their matching segmentation masks $\tilde{S}_{A}$, $\tilde{S}_{C}$ and $\tilde{S}_{S}$ are produced by corresponding decoders, decoder $A$, decoder $C$ and decoder $S$. }
\State Freeze only the parameters of the encoder, decoders $S$ and $C$; then, train the decoder $A$ with the loss $L_{V=A}^{d}$, in Eq. (\ref{eq:LossFunc2}) 
\State Freeze only the parameters of the encoder and decoders $A$ and $S$; then, train the decoder $C$ with the loss $L_{V=C}^{d}$, in Eq. (\ref{eq:LossFunc2})
\State Freeze only the parameters of the encoder and decoders $A$ and $C$; then, train the decoder $S$ with the loss $L_{V=S}^{d}$, in Eq. (\ref{eq:LossFunc2})
\EndFor
\end{algorithmic}
\end{algorithm}
\footnotetext{Although training the encoder with varying numbers of input images $I_{A}$, $I_{C}$ and $I_{S}$ may bias the encoder towards the most frequently seen view images, we leave the investigation of this problem for future work.}

\subsection{Run-time Quality Control}  \label{sec:Method_MultipleAttackDetection}

We expect that a multi-view network, trained similar samples to a volume image, to produce very similar segmentation masks for its resampled versions at three views. To be more precise, it should satisfy $G(f_{A}(I_{A})) \approx G(f_{S}(I_{S})) \approx G(f_{C}(I_{C})) $, where $f_{A}$, $f_{S}$ and $f_{C}$ respectively correspond to decoder $A$, decoder $S$ and decoder $C$ in Figure \ref{fig:Pipeline}. $G(\cdot)$ transforms back decoder outputs to a preselected view. Because of averaging decoder outputs to generate a final segmentation mask, we hypothesise that segmentation performance relates to maximum similarity between decoder outputs. To quantify this agreement, we propose a max consistency score $max\mathcal{C}_{S,A,C}$ with Eq. (\ref{eqn:CosSim}), which is based on pairwise cosine similarities of decoders' output volumes.

\begin{align} 
&max\mathcal{C}_{S,A,C} =max(\mathcal{C}_{S,A},\mathcal{C}_{S,C},\mathcal{C}_{C,A}) \label{eqn:CosSim}\\
&\mathcal{C}_{a,b}=\frac{ G^{c}(V_{a})\cdot  G^{c}(V_{b}) }{|| G^{c}(V_{a}) || \cdot || G^{c}(V_{b}) ||} , \quad -1< \mathcal{C}_{a,b} <1  \quad \& \quad a\neq b  \quad \&  \quad  V_{a}, V_{b} \in \{ V_{S},V_{A},V_{C}\} 
\end{align}

where $max\mathcal{C}_{S,A,C}$ is equal to the maximum of three cosine similarities calculated for three different permutations of decoder pairs: sagittal- axial, sagittal-coronal and coronal-axial. $\mathcal{C}_{a,b}$ is a scalar in the range of $[-1,1]$, showing the cosine similarity between $G^{c}(V_{a}) $ and $G^{c}(V_{b})$, where $G^{c}(\cdot)$ converts standardised outputs with the function $G(\cdot)$ to column vectors, $ G^{c}( V_{a} )\in R^{NWHx1}$ and $ G^{c}(V_{b}) \in R^{NWHx1}$. Because negative correlation between decoder outputs is not meaningful for robustness evaluation, large positive values of $\mathcal{C}_{a,b}$ show good agreement between decoder outputs.

We examine the relation between $max\mathcal{C}_{S,A,C}$ and segmentation performance measured with Jaccard indices, which shows a larger Pearson correlation coefficients obtained with Dice scores (see Table \ref{table:Pearson2013} - \ref{table:Pearson2018}). For this experiment, we use the original MRI volumes along with their corrupted versions at various noise magnitudes with Rician or engineered noise to mimic samples likely to lead to robustness issues, which reduces the trust of medical clinicians to segmentation results. Therefore, we are able to assess if the run-time Jaccard score of a segmentation mask is larger than a predetermined threshold by only examining its $max\mathcal{C}_{S,A,C}$, without comparing it with its ground truth mask or without knowing the similarity of an input image to training set.

Low agreement between decoder outputs can reveal which view images can reduce segmentation accuracy. In this case, we take the minimum of cosine similarities, \textit{min consistency score} $min\mathcal{C}_{S,A,C}=min(\mathcal{C}_{S,A},\mathcal{C}_{S,C},\mathcal{C}_{C,A})$, to detect the worst segmented view. This score will be used in Section \ref{sec:Ablation} to explain why TMS-Net is robust to noisy images.

\subsection{Degraded Segmentation Mask Generation} 

It is important to generate various qualities of segmentation masks to assess the performance of a run-time quality estimation method. Here, we consider three potential reasons for poor segmentation masks: (i) undertrained networks or low capacity networks, expected to show poor segmentation performance for each view of a volume image, (ii) high anisotropic resolution, the number of slices sampled from different views may vary, which may, in turn, lead to large performance differences across decoders of a multi-view network and (iii) inappropriate imaging settings, where the visibility of the organ largely decreases.

In contrast to previous studies training several models \cite{hann2021ensemble,robinson2019automated,valindria2017reverse} to produce a range of poor segmentation masks, we corrupt images with engineered and Rician noise; this allow us to simulate the aforementioned problems. We simulate the first two scenarios by corrupting images with engineered noise on all three views or a single view consecutively. To emphasise how many views are corrupted, we call them \textit{three views corruption} and \textit{single view corruption}. For three views corruption, noise is added to each slice of a volume image  after reslicing the original image for each view. For the single view corruption, after corrupting a selected view, the noisy volume is resliced to generate images for the other views.

Although engineered noise degrades segmentation quality, it preserves the visibility of anatomical structures in the input images. To reduce their visibility, we add moderate-to-severe Rician noise to images to simulate the last problem. We also adjust the quality of segmentation masks by varying noise magnitude and also by using two types of engineered noise; one is stronger. This corresponds to changing model capacity in the first problem, for example using smaller version of a model, and varying training data size for a specific view in the second problem, due to the amount of anisotropy.

\subsubsection{Engineered Noise}

The outputs of deep networks can be dramatically changed by manipulating input images by adding invisible but carefully-engineered noise to them, so that the networks produce irrelevant results to their purpose \cite{krittanawong2019deep,lu2021clinical}. In this paper, we use the fast gradient sign method (FGSM) and basic iterative method (BIM), which will be explained as follows \cite{goodfellow2014explaining,kurakin2016adversarial}, for robustness evaluation because of their stronger ability to reveal any weakness of a trained model.

FGSM uses the signs of the gradients of the loss with respect to input images to generate engineered perturbations. A perturbed image $I_{adv}$ with a noise strength of $\epsilon$ can be generated from the original image $I$ and its segmentation mask $S$, with the formula of $I_{p}=I+\epsilon . \ sign \big(\Delta_{ I} L \big(f(I, \theta),S \big) \big)$, where $f$ denotes a network parametrised by $\theta$ and $L$ is a loss function. Kurakin \textit{et al.} extended FGSM method by iteratively updating perturbed images with $I_{p}^{n+1}= C\big\{I_{p}^{n}+\alpha . \ sign \big(\Delta_{I} L \big(f(I_{p}^{n}, \theta),S \big) \big)\big\}$, where  $ I_{p}^{0}=I$ and $C(\cdot)$ clips the intensities of perturbed images to keep them in the range of the original intensities \cite{kurakin2016adversarial}. This method is called BIM \cite{kurakin2016adversarial}. 

\subsubsection{Rician Noise } \label{sec:noiseDis}
Magnitude MR images that do not employ multi-coil parallel reconstruction methods have a Rician noise distribution. A noisy image $I_{N}$ can be represented with $I_{N}=\sqrt{(I_{C}+G_{1})^{2}+G_{2}^{2}}$, where $I_{C}$ is a noise-free magnitude MRI image and $G_{1}$ and $G_{2}$ are the experimental Gaussian noise with a standard deviation of $\sigma_{1}=\sigma_{2}=\sigma$ \cite{mohan2014survey}. Rician noise has a signal-dependency property that the noise comes closer to a Rayleigh distribution for low signal-to-noise ratio (SNR) images though it shows a Gaussian distribution for high SNR images \cite{mohan2014survey}.

\subsection{Ablation Studies} \label{sec:Ablation}

We conduct ablation experiments to see if constraining latent spaces of three different orthogonal view inputs improves network's robustness. For this purpose, we trained a modified version of TMS-Net, without parameter sharing, where three encoders are connected to three decoders similar to training three independent networks and compare its performance with that of the original design of TMS-Net. In addition to examining overall performance, we use min consistency scores to show how much a poor performing view can affect Jaccard indices. In these experiments, we corrupt input images with three views engineered noise and Rician noise; the latter is applied to the standard view. We also examine T-SNE plots of the last encoder layer activations of TMS-Net and TMS-Net $3$ for an MRI volume from STACOM $2013$ to provide interpretation of working mechanism of the networks.

\section{Experimental Setting}\label{sec:ExperimentalSetting}
\subsection{Material}
We assessed the segmentation performance of TMS-Net on STACOM $2013$ \cite{tobon2015benchmark} \footnote{\url{http://www.cardiacatlas.org/challenges/left-atrium-segmentation-challenge/} } and STACOM $2018$ \cite{xiong2021global} \footnote{\url{http://atriaseg2018.cardiacatlas.org} } challenge datasets. STACOM $2013$ dataset consists of $30$ balanced steady state free precession (bSSFP) MRI images with a resolution of $1.25~mm ~\times 1.25 ~mm ~ \times 2.7 ~mm$, where $10/20$ images were given for training/test by the challenge organisers. The dataset has a mixture of high, moderate and poor quality images. The ground truth images include labels for LA body, a part of pulmonary veins and appendage; though, the appendage was not labelled for the test set so its is not included in performance evaluation by the organisers. Because we want to assess noise robustness of TMS-Net, we unified the labels for LA body and pulmonary veins, by assigning the same label to both structures.  

The STACOM $2018$ dataset contains $154$ late Gadolinium-enhanced (LGE) MRI volumes, collected from AF patients, with an isotropic resolution of $0.625~mm \times ~ 0.625 ~mm \times ~ 0.625~mm$. Of these, $100$ MRI volumes are publicly available and released as the training dataset by the challenge organisers. We allocated the first $70$ volumes of the $100$ volumes for training and the last $20$ for test and the remained $10$ for validation. The dataset has a fixed number of slices of $88$. To reduce training time, we downsample the images by a factor of $2$ at the $x-y$ plane. The dataset contains images from multiple medical centers. The segmentation masks include LA body, the mitral valve, LA appandage and some parts of pulmonary veins.  The majority of images in this dataset is medium quality and there is equal number of low and high quality images.

\subsection{Baseline Networks}
We compare the segmentation and robustness performance of TMS-Net with that of U-Net and SegNet, which are well-known networks for image segmentation \cite{ronneberger2015u, badrinarayanan2017segnet} and recently proposed networks, CE-Net \cite{gu2019net} and U-Net $3+$ \cite{zhou2018unet++}. Because ensemble models generally outperform single models, we also train the latter networks in multi-view  training settings, where we respectively call them U-Net-M, SegNet-M, CE-Net-M and U-Net $3+$ -M to indicate this modification. For these networks, the techniques explained in Section \ref{sec:TMS-Net} are also used to combine segmentation maps from the three views after training an independent network for each of the sagittal, axial and coronal view. To avoid overfitting in U-Net and SegNet and their multi-view versions, we integrated dropout layers with a probability of $0.5$ into them, which were added after the eighth and tenth convolutional layers of the U-Net and between the fifth and sixteenth convolutional layers of the SegNet. We trained the U-Net, the SegNet, CE-Net and U-Net$3+$ with the predetermined standard view of each dataset. We used the last layer's loss for engineered noise generation for Unet$3+$.

\subsection{Network Training}
We optimised the parameters of all networks with the Adam algorithm and a learning rate of $0.0005$, which is decreased with an exponential decay learning rate of $0.9$ at each epoch. Then, we manually tuned the weight decay parameter per network, which was $0.01$ for SegNet and TMS-Net and $0.005$ for U-Net for STACOM $2013$ dataset and $0.001$ for the all networks for STACOM $2018$ dataset. Training lasted for approximately $10$ epochs for TMS-Net and between $60$ and $70$ epochs for U-Net and SegNet and we manually stopped it when any sign of overfitting was observed. The batch size in the training set was $4$ for STACOM $2013$ dataset and $8$ for STACOM $2018$, with an image size of $192\times 192$ and $128 \times 128$ pixels respectively, which are cropped around the image center.

To avoid overfitting to the training dataset, we performed on-the-fly data augmentation, which included translation up to $\left[ \pm 30,\pm 30\right]$ pixels from slice center, rotation by up to $\pm 45^{o}$, scaling in the range of $\left[0.8,1.2\right]$, shear transformation up to $\left[\pm 10, \pm 10\right]$ and contrast augmentation in the range of $\left[0.8, 1.2\right]$. To increase the number of LA slices relative to that of non-LA slices in training datasets, we sampled non-LA volume with $5$ slices' intervals, by excluding the first five closest slices to the LA volume at both ends. We used a Pytorch Wavelets library \cite{cotter2020uses} \footnote{\url{https://pytorch-wavelets.readthedocs.io/en/latest/index.html}}, for the wavelet analysis. All networks were trained with Pytorch, whose version is $1.10.2$, on a computer with two Geforce GTX $1080$ GPUs.

\subsection{Generating Corrupted Images}
We normalise the intensity distribution of each slice in the range of $\left( 0,1\right)$. For BIM engineered noise, iteration number and $\alpha$ respectively were set to be $5$ and $1$. For all corrupted images, after adding the noise to input images, we clamped intensities between zero and one. We determine the sagittal view for STACOM $2013$ and the axial view for STACOM $2018$ as the standard view and use a $2D$ Gaussian kernel with a radius of $2$ for smoothing of transformed segmentation masks. Noise is added to the the standard view for single view corruption.

\subsection{Performance Metrics}
The segmentation performance is assessed with Dice and Jaccard scores, calculated with a Python library called MedPy \footnote{https://pypi.org/project/MedPy/}, after thresholding network outputs with Otsu's method. We report Pearson correlation coefficient to quantify relation between our max consistency scores and Jaccard indices. We also give mean absolute error (MAE) and area under the curve (AUC) of receiver operating characteristic curve (ROC) to assess performance for binary run-time quality estimation.

\section{Results} \label{sec:Results}
\subsection{Ablation Studies} \label{sec:AblationExp}

\begin{figure}[!h]
\centering
\includegraphics[width=0.8\textwidth]{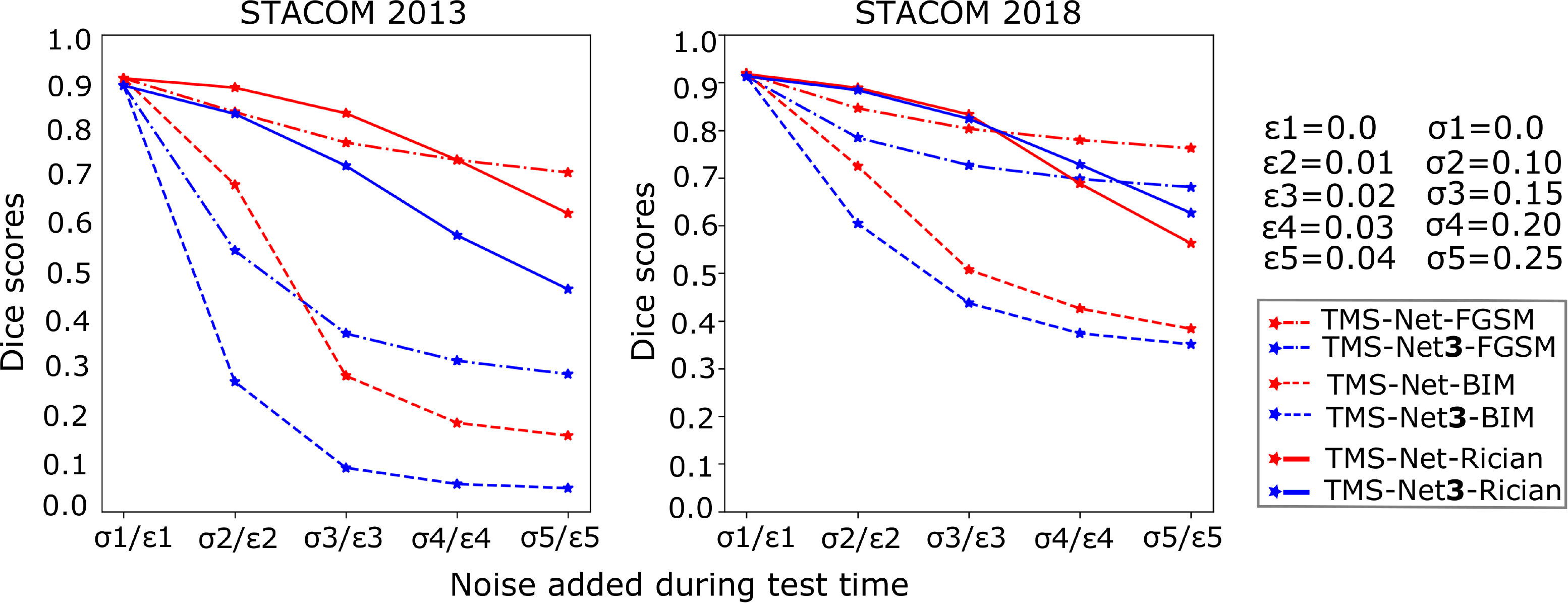}
\caption{Results of ablation experiments on STACOM $2013$ and STACOM $2018$ datasets. TMS-Net shows more robustness to engineered and Rician noise by outperforming TMS-Net$3$ for the same amount of noise. (Best viewed in color.) } \label{fig:FGSM3attackAblation}
\end{figure}

We find that the use of shared encoder in the design of TMS-Net plays an important role in its noise robustness, regardless of noise types (See Figure \ref{fig:FGSM3attackAblation}). Particularly, we observe that using independent encoders in TMS-Net leads to more performance drop on STACOM $2013$ compared to STACOM $2018$ dataset, which may be explained with the limited size of training set for STACOM $2013$. Given larger gaps between the performance of TMS-Net and that of TMS$3$-Net in STACOM $2013$ compared to STACOM $2018$, we can conclude that TMS-Net is less prone to overfitting in the presence of a small size of training set.

\begin{figure}[!h]
\centering
\includegraphics[width=0.9 \textwidth]{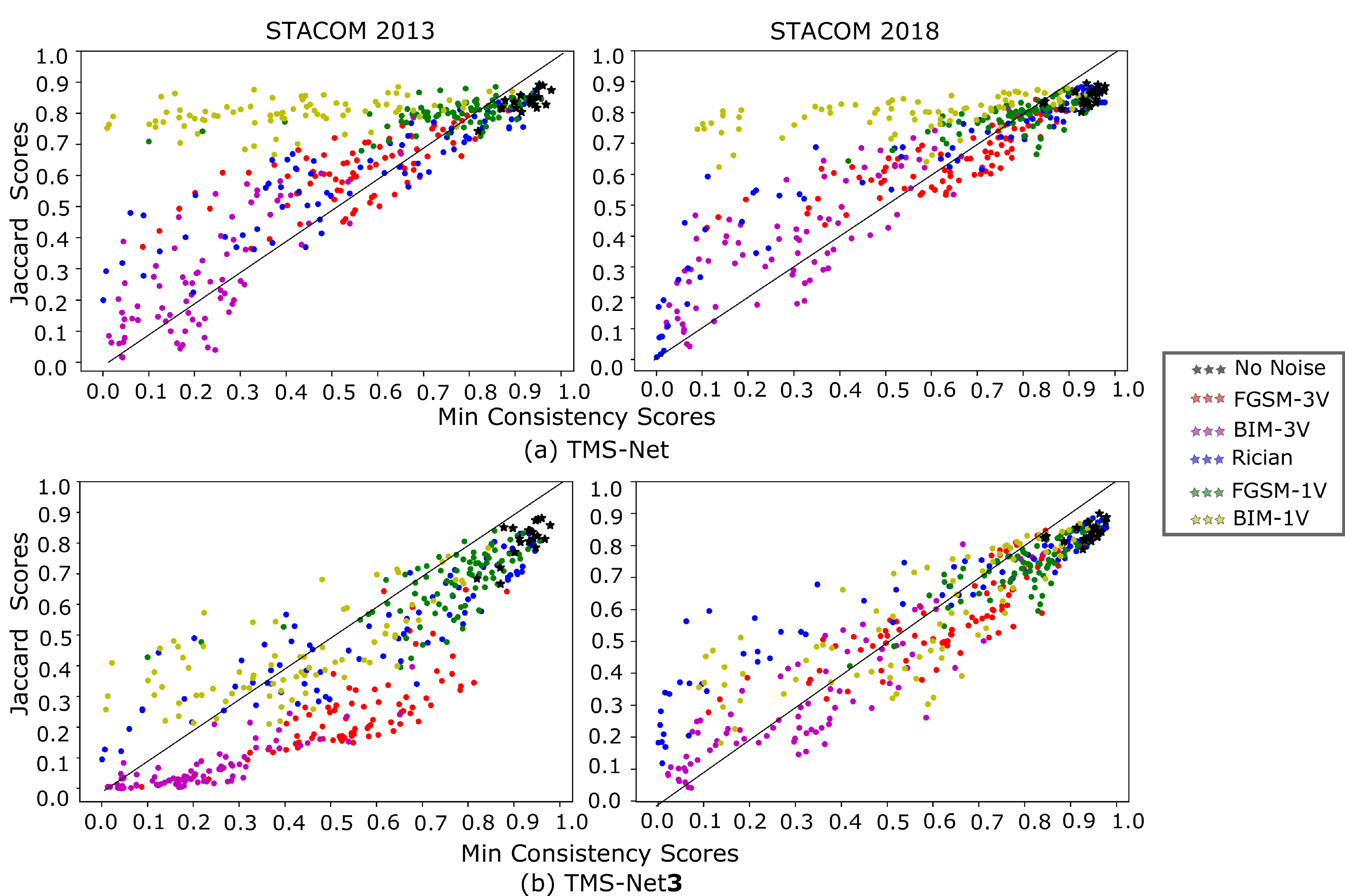}
\caption{Min consistency scores vs Jaccard indices. Black dots represent results for the original images and the others are noisy images. See legends for which corruption method is used for which images. Dots over the diagonal lines show results of image samples for which disagreement across decoder outputs are better tolerated, by indicating robust segmentation. Note that high robustness of TMS-Net for single view corruption --FGSM-$1V$ and BIM-$1V$-- , where the network preserves high Jaccard indices, regardless of how different the targeted view is segmented than other views. (Best viewed in color.)} \label{fig:DistributionMinConsistecy}
\end{figure}

To further investigate the effect of sharing encoder on the noise robustness of TMS-Net, Figure \ref{fig:DistributionMinConsistecy}(a) shows the relation between min consistency scores across its decoder outputs and Jaccard indices of mean decoder outputs. In this figure, even though some samples have at least one view with very low similarities to others, indicated with small \textit{min consistency scores} of $[0,0.5]$, they have large Jaccard scores, mostly larger than $0.7$, particularly true for images corrupted with BIM-$1V$ noise. This explains why TMS-Net is more robust than TMS-Net$3$, which is because of better compensating missed or corrupted information. On the other hand, we observe almost a linear relation between \textit{min consistency scores} and Jaccard indices, for TMS-Net$3$ in Figure \ref{fig:DistributionMinConsistecy}(b), which suggests that the network is sensitive to issues related to the noise types such as high anisotropy and small training set.

\begin{figure}[!h]
\centering
\includegraphics[width=0.5 \textwidth]{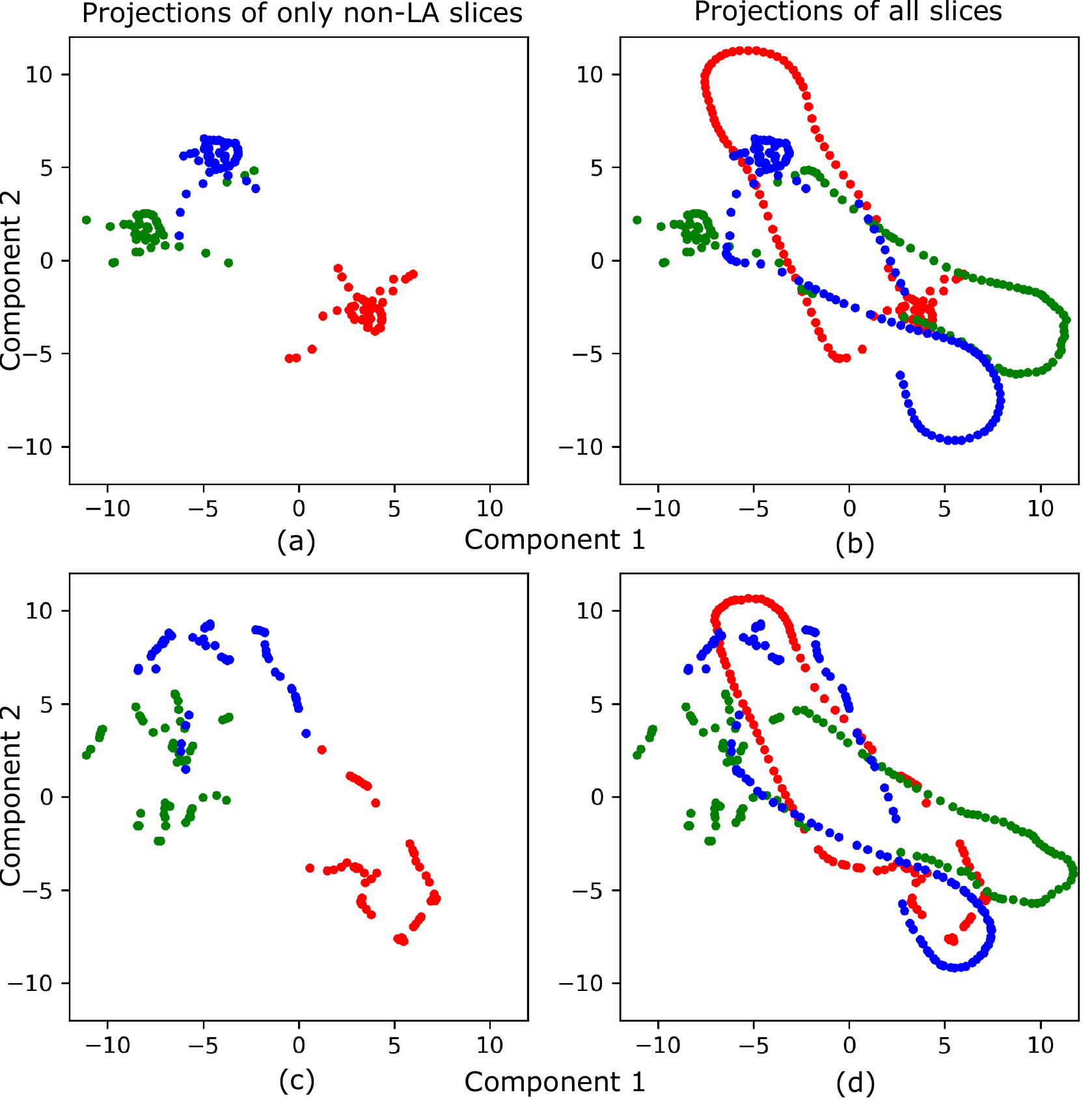}
\caption{Typical t-SNE plots for the last encoder layer activations of TMS-Net in (a)-(b) and that of TMS-Net$3$ in (c)-(d). Right column images (a) and (c) show only non-LA slices' projections while left column images (b) and (c) display all slices' projections. Red, green and blue dots respectively denote $2D$ projections for coronal, sagittal and axial slices of the same MRI volume. In contrast to TMS-Net$3$, TMS-Net have different representations for non-LA and LA slices, the former forms clusters while the latter generates $U$ or $8$ like curves. This shows the ability of TMS-Net to effectively discriminate LA slices from non-LA slices. (Best viewed in color.)} \label{fig:tSNEPlot}
\end{figure}

\begin{figure}[!h]
\centering
\includegraphics[width=0.99\textwidth]{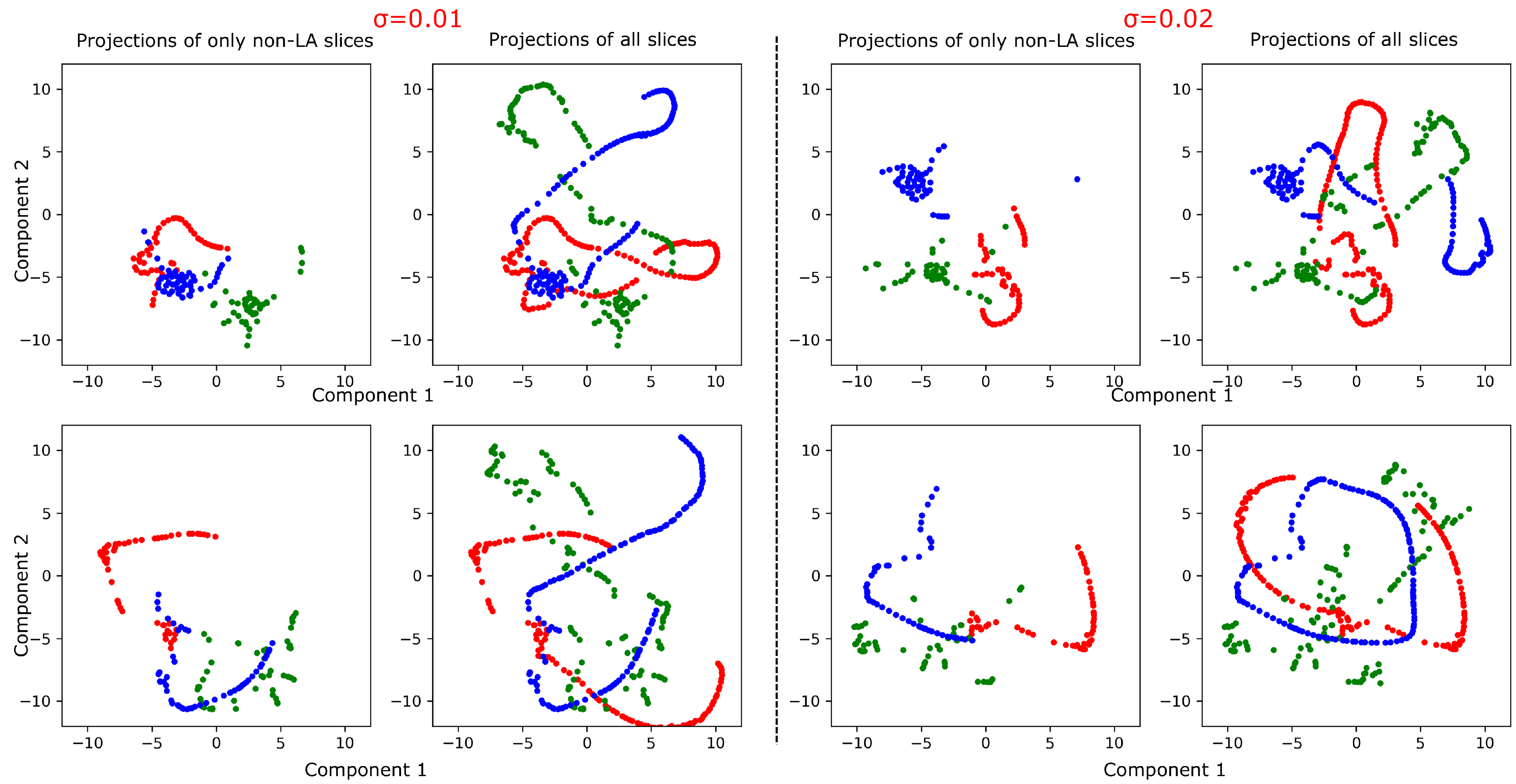}
\caption{T-SNE plots for the last encoder layer activations of TMS-Net (the top row) and that of TMS-Net$3$ (the bottom row) after corrupting input images with three views corruption of FGSM of $\sigma=0.01$ and $\sigma=0.02$. See Figure \ref{fig:tSNEPlot} for more details. (Best viewed in color.)} \label{fig:tSNEPlot2}
\end{figure}

Figure \ref{fig:tSNEPlot} compares t-SNE plots of lateral representation of TMS-Net and TMS-Net$3$ for an MRI volume from STACOM $2013$. In contrast to TMS-Net$3$, non-LA slice projections are clustered around three centers for TMS-Net; each corresponding to a different view. This means that the shared encoder of TMS-Net can accurately discriminate non-LA slices from LA slices in its last encoder layer and treat them differently, in contrast to TMS-Net$3$ which generated similar response to LA slices. Another observation from the figure is that the projections of LA slices form $U$ or $8$-like curves for both networks, by representing regular changes in LA shape across slices such as typical decreases of LA size toward end slices. When the input image volume is corrupted by three views corruption of FGSM, the regular curves of the projections for the original images are changed, as shown in Figure \ref{fig:tSNEPlot2}, such that, some non-LA slices are identified as the LA, indicated by small curves in non-LA slice projections. Also, we observe small perturbations in LA slice projections, deviating from their regular shapes, which can be as a result of 
changes in network responses to some LA slices due to the corruption of input images with the engineered noise. These alterations are more prominent for TMS-Net$3$ supporting superiority of TMS-Net on noise-robustness. As a result, we can conclude that t-SNE plots of the last encoder layer activations of TMS-Net have human interpretable features, which is valuable to improve trust in network outputs.

\subsection{Degraded Images For Poor Segmentation Mask Generation}
\begin{figure*}[!h]
\centering
\includegraphics[width=0.99 \textwidth]{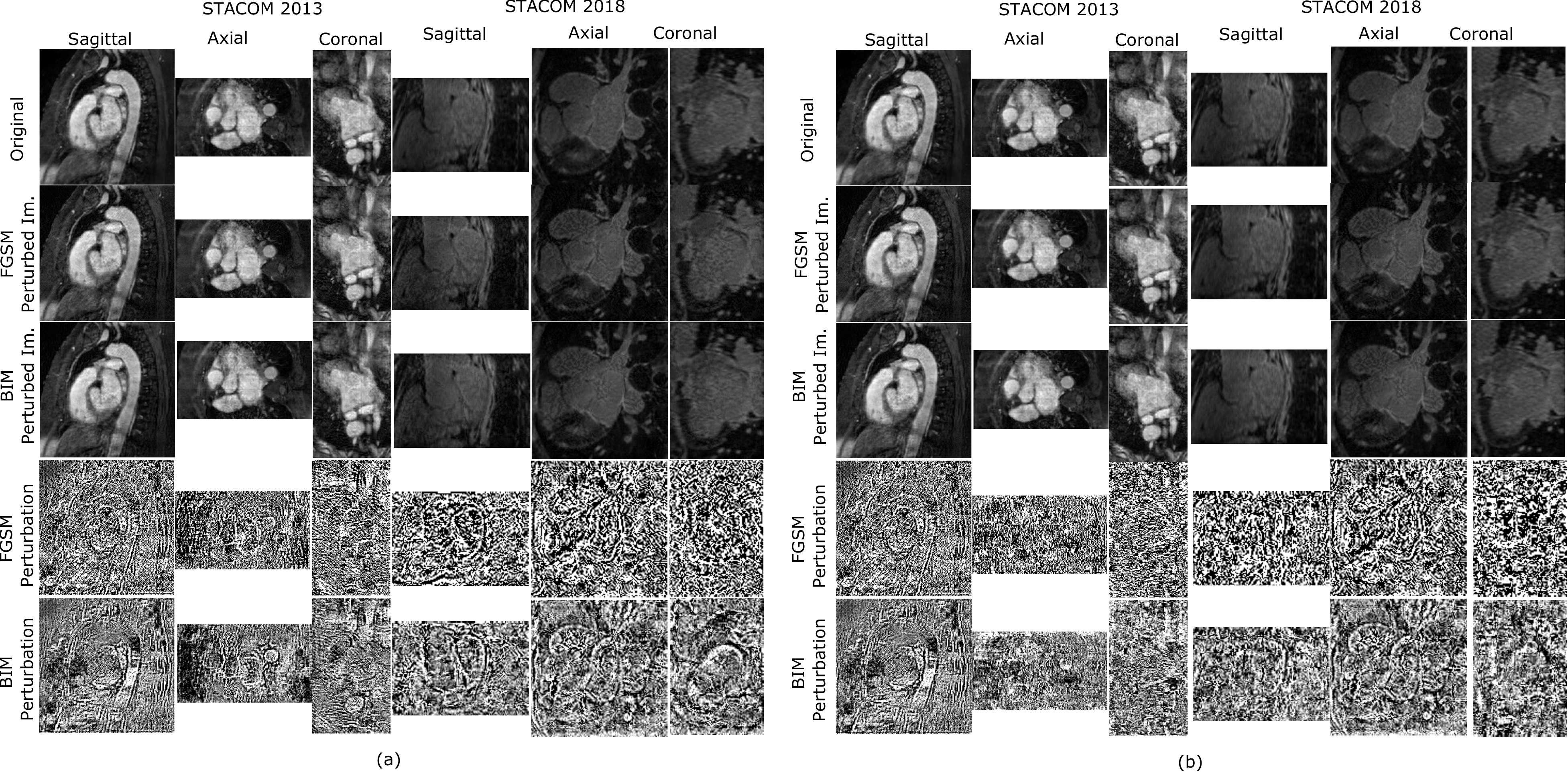}
\caption{Typical examples of noisy images and perturbation itself for (a) three views corruption (b) single view corruption, generated for TMS-Net by FGSM and BIM, on STACOM $2013$ and STACOM $2018$ datasets. Note that noise is not visible in noisy (perturbed) images due to its small magnitude, $\epsilon=0.03$; however, it is effective to reduce segmentation quality due to exploiting weakness of the network for a targeted view/s (see Figure \ref{fig:ExampleOutputs} in Appendix \ref{sec:3viewAdvers}). Perturbation images for three views corruption mostly mimic high frequency features of the original views. On the other hand, for single view corruption, only targeted views -- the sagittal for STACOM $2013$ and the axial for STACOM $2018$ -- has structural similarities to the original ones, where the same noise behaves as random noise when transformed to other views. (Images from STACOM $2018$ are resized to $192 \times 192$ for better visualisation. Perturbation images are magnified by roughly $\times 35$.) }\label{fig:SAC}
\end{figure*}

\begin{figure}[!h]
\centering
\includegraphics[width=0.65 \textwidth]{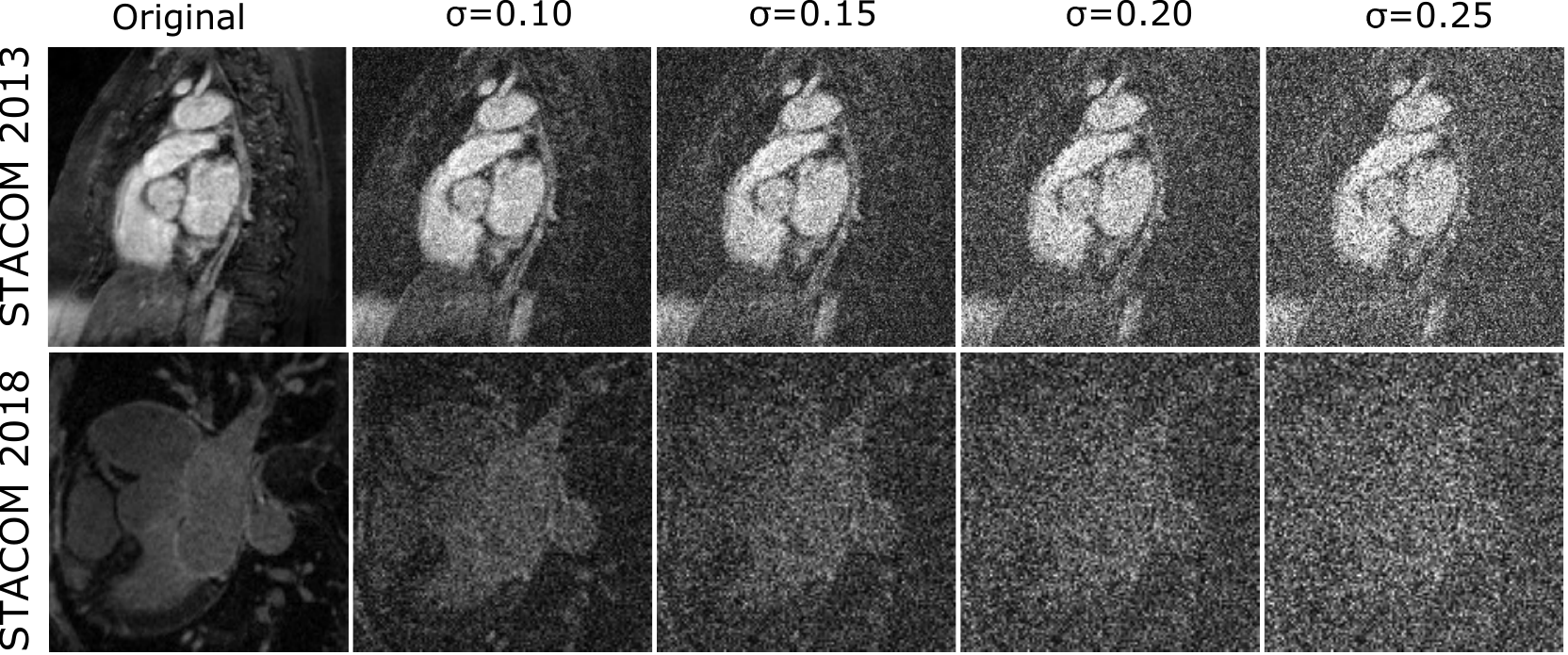}
\caption{Corrupted images with various variances ($\sigma$) of Rician noise from STACOM $2013$ and STACOM $2018$ datasets.} \label{fig:NoiseFree2013_2018}
\end{figure}

Figure \ref{fig:SAC} exemplifies corrupted images by FGSM and BIM methods when a single or three views are corrupted by comparing them with the original images. As understood from the figure, there is no visible difference apparent between the original and corrupted images, which exhibit big potentials for simulating undertraining and anisotropy related problems. On the other hand, Figure \ref{fig:NoiseFree2013_2018} demonstrates images corrupted with synthetic Rician noise, where the visibility of anatomical structures is largely reduced, simulating poor imaging settings. See Figure \ref{fig:ExampleOutputs} in Appendix for how much consistency decreases between decoder outputs in the presence of three views engineered noise and Rician noise.

\subsection{Segmentation Performance and Robustness Comparison} \label{sec:Perform}

\begin{figure*}[!h]
\centering
\includegraphics[width=0.6 \textwidth]{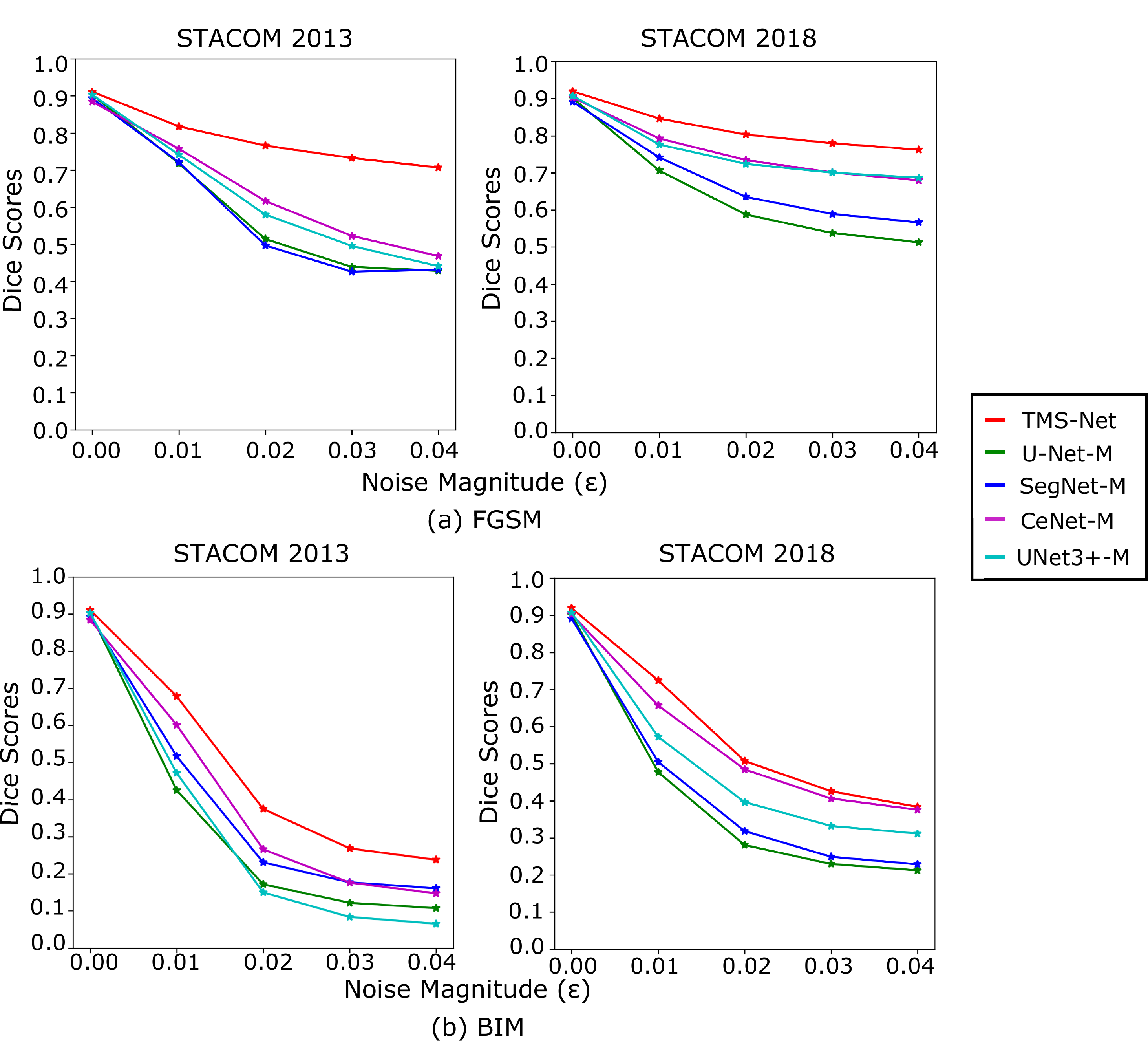}
\caption{Robustness to three views engineered corruption. Note far smaller performance decrease in TMS-Net in the presence of FGSM noise compared to the other multi-view networks. Similar performance is also displayed for BIM on STACOM $2018$, which is much stronger than FGSM. (Best viewed in color.)}\label{fig:FGSMattackMultipleViews} 
\end{figure*}

\begin{figure*}[!h]
\centering
\includegraphics[width=0.6\textwidth]{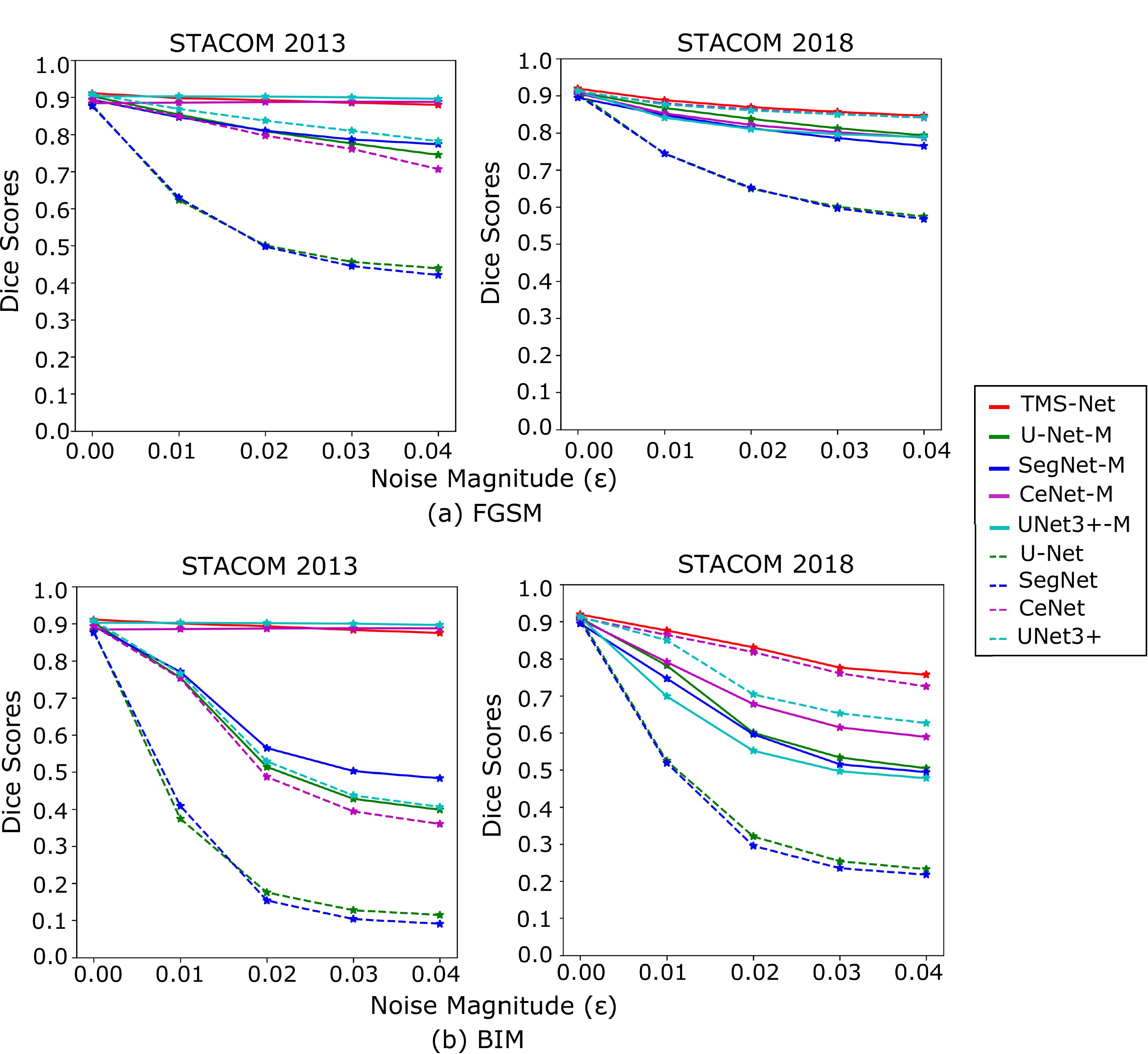}
\caption{Robustness to single view engineered corruption. Note that information loss/corruption in a single view does not affect TMS-Net as much as the other networks, even for strong BIM noise. (Best viewed in color.)} \label{fig:FGSMattackSingleViews} 
\end{figure*}

\begin{figure}[!h]
\centering
\includegraphics[width=0.6 \textwidth]{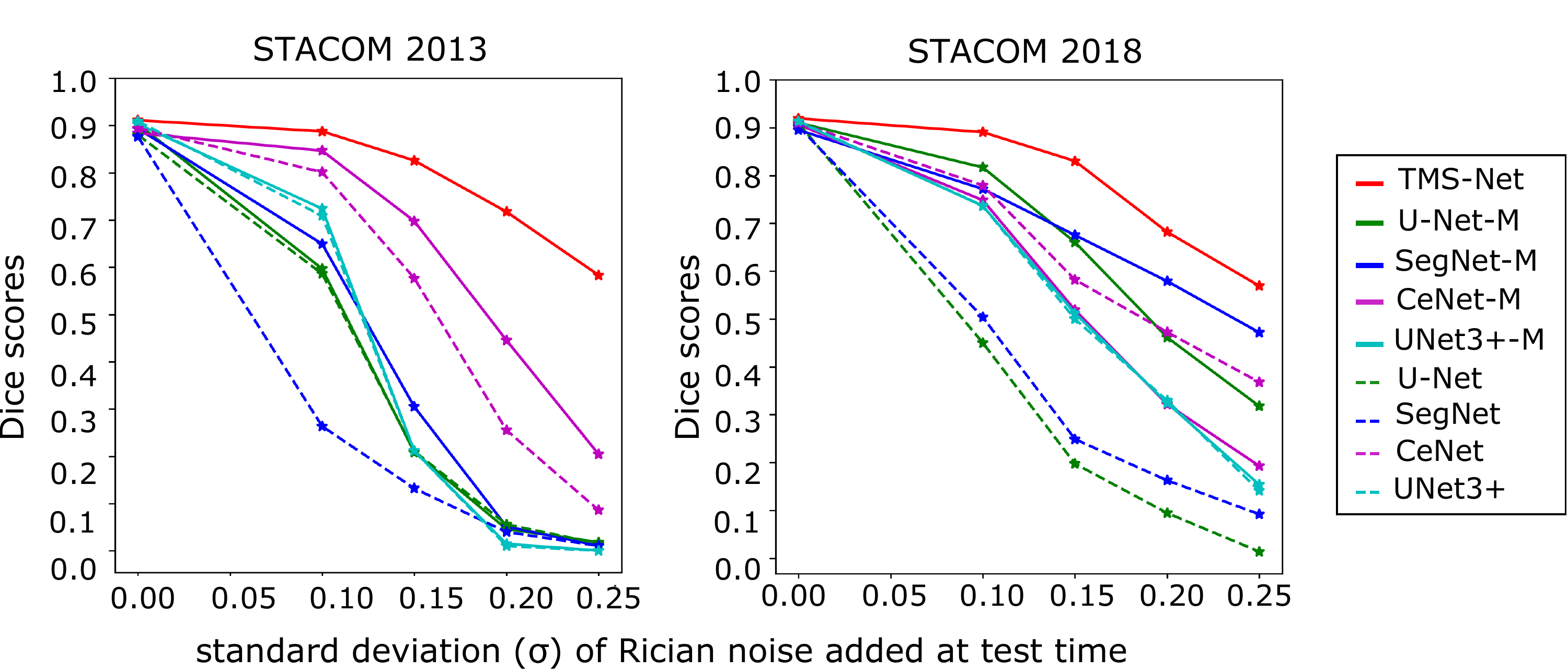}
\caption{Robustness to additive Rician noise. TMS-Net presents large Dice scores of over $0.6$ even when the other networks produce Dice scores of around $0.0$ for STACOM $2013$, at $\sigma=0.25$. Also, note the insensitivity of the performance of TMS-Net to different datasets. (Best viewed in color.) } \label{fig:Graphs_Rician} 
\end{figure}

Table \ref{table:Noise_Free_Results_2013} and Table \ref{table:Noise_Free_Results_2018} tabulate the performance of TMS-Net and previous methods on the original images of STACOM $2013$ and STACOM $2018$ datasets. Our network produces better or comparable performance to previous methods. We compare the robustness of TMS-Net with other networks in Figure \ref{fig:FGSMattackMultipleViews} for three views corruption with engineered noise, in Figure \ref{fig:FGSMattackSingleViews} for single view corruption with engineered noise and in Figure \ref{fig:Graphs_Rician} for Rician noise, respectively simulating undertrainig, high anisotropy and poor image settings problems. As understood from the figures, TMS-Net is the most robust network to the noise types, followed by other multi-view networks. We have two conclusions from these results. The first, TMS-Net can better deal with corrupted information at not only single view but all three views, indicating more tolerance to high anisotropy and undertraining related problems. Also, the network is more robust to poor imaging settings by mostly maintaining its segmentation performance for severe Rician noise, where the other networks generate almost zero Dice scores  (see Figure \ref{fig:Graphs_Rician}). Secondly, compared to single networks, not only TMS-Net but also other multi-view networks are more robust to challenging conditions.

\begin{table}[!h]

\centering
\caption{Performance comparison on the original Stacom $2013$.  \label{table:Noise_Free_Results_2013}}
\resizebox{0.5 \textwidth}{!}{%
\begin{threeparttable}
\begin{tabular} {@{} lc cc@{} } \toprule %
Method&Dice & Jaccard & Parameter Count\\ 
\cmidrule(lr){1-1} \cmidrule(lr){2-2}  \cmidrule(lr){3-3} \cmidrule(lr){4-4} 

TMS-Net& \textbf{0.91} $\pm$ {0.02} & \textbf{0.83} $\pm$ {0.04} & 8 million\\  

U-Net & 0.88 $\pm$ 0.05 & 0.79 $\pm$ 0.07& 17.3 million  \\ 
SegNet&  0.88 $\pm$ 0.03 & 0.78 $\pm$ 0.05&  15.3 million \\

U-Net-M  &0.90 $\pm$ 0.03 & 0.82 $\pm$ 0.04 & 3 $\times $ 17.3 million \\ 

SegNet-M  &0.88 $\pm$ 0.03 & 0.79 $\pm$ 0.05 &  3 $\times $ 15.3 million \\ 
U-Net $3+$ \cite{zhou2018unet++} & \textbf{0.91} $\pm$ 0.02 & \textbf{0.83} $\pm$ 0.04 &17.3 million\\
U-Net $3+$-M \cite{zhou2018unet++} & 0.90 $\pm$ 0.03 & 0.82 $\pm$ 0.04 &3 $\times $ 17.3 million\\
CE-Net \cite{gu2019net}& 0.89 $\pm$ 0.03 & 0.81 $\pm$ 0.05 &17.3 million\\
CE-Net-M \cite{gu2019net}& 0.88 $\pm$ 0.03 & 0.79 $\pm$ 0.05 &3 $\times $ 17.3 million\\
\midrule

CardiacNet \cite{mortazi2017cardiacnet} & 0.91 \\
LTSI-VRG \tnote{1} & 0.86\\ 
UCL-1C \tnote{1} & 0.89\\ 
OBS-2 \tnote{1} & 0.91\\

\bottomrule
\end{tabular}
    \begin{tablenotes}\footnotesize
        \item[1] Results were taken from \cite{mortazi2017cardiacnet}.
    \end{tablenotes}

\end{threeparttable}
}

\end{table}

\begin{table}[!h]

\centering
\caption{Performance comparison on the original Stacom $2018$.  \label{table:Noise_Free_Results_2018}}
\resizebox{0.5 \textwidth}{!}{%
\begin{threeparttable}
\begin{tabular} {@{} lc cc@{} } \toprule %
Method&Dice & Jaccard & Parameter Count\\
\cmidrule(lr){1-1} \cmidrule(lr){2-2}  \cmidrule(lr){3-3} \cmidrule(lr){4-4} 
   
TMS-Net &\textbf{ 0.92} $\pm$ {0.02} & \textbf{0.85} $\pm$ {0.03}  & 8 million\\

U-Net  &0.90 $\pm$ 0.02 & 0.82 $\pm$ 0.04 &  17.3 million \\ 

SegNet  &0.88 $\pm$ 0.03 & 0.78 $\pm$ 0.05 &  15.3 million \\

U-Net-M  &0.91 $\pm$ 0.02 & 0.84 $\pm$ 0.03 & 3 $\times $ 17.3 million \\

SegNet-M  &0.89 $\pm$ 0.02 & 0.81 $\pm$ 0.04  & 3 $\times $ 15.3 million \\ 
U-Net $3+$ \cite{zhou2018unet++} & 0.91 $\pm$ 0.02& 0.84 $\pm$ 0.03 &17.3 million\\
U-Net $3+$-M \cite{zhou2018unet++}& 0.91 $\pm$ 0.02& 0.84 $\pm$ 0.03 &3 $\times $ 17.3 million\\
CE-Net \cite{gu2019net}&0.91 $\pm$ 0.02& 0.84 $\pm$ 0.03&17.3 million\\
CE-Net-M \cite{gu2019net}& 0.91 $\pm$ 0.02 & 0.83 $\pm$ 0.03 &3 $\times $ 17.3 million\\

\midrule

Chen \textit{et al.} \cite{chen2018multi} & 0.90 $\pm$ 0.03& 0.82 $\pm$ 0.06 \\ 

Yang \textit{et al.} \cite{yang2018combating}&0.92 &0.86& \\ 
Jia \textit{et al.} \cite{jia2018automatically}  &0.92 $\pm$ 0.03& \\ 
Li \textit{et al.} \cite{li2018attention}&0.92&\\ 

\bottomrule
\end{tabular}

\end{threeparttable}
}
\end{table}

\subsection{Run-time Performance Estimation} \label{sec:Attack}

\begin{figure}[!h]
\centering
\includegraphics[width=0.9 \textwidth]{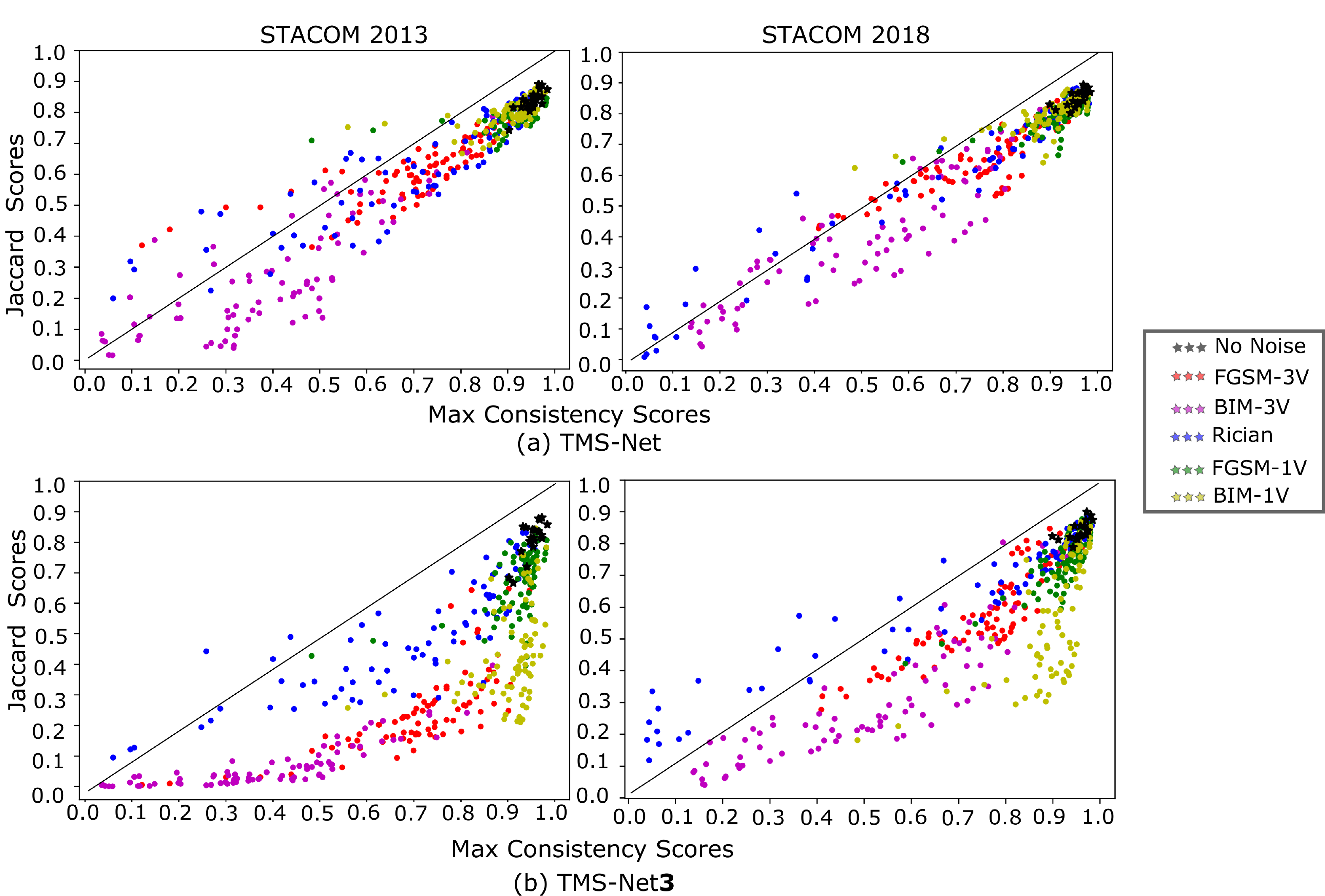}
\caption{Max consistency scores vs Jaccard indices. Ideally, dots should be located on the diagonal lines for unbiased estimation of Jaccard indices at run-time. Note the linearity of the relation between max consistency scores and Jaccard indices for TMS-Net. (Best viewed in color.)} \label{fig:DistributionMaxConsistecy}
\end{figure}

Figure \ref{fig:DistributionMaxConsistecy} demonstrates relation between our max consistency scores and run-time Jaccard scores, calculated from ground truth segmentation masks. As shown in the figure, TMS-Net presents a linear relation between these parameters, which is degraded when parameter sharing is removed in TMS-Net$3$. Pearson correlation coefficents are given in Table \ref{table:Pearson2013} for STACOM $2013$ and Table \ref{table:Pearson2018} for STACOM $2018$, where we find a very high correlation between our max consistency scores and Jaccard indices. Although our method is not based on training a model for run-time performance estimation, we obtain small MAEs as presented in Table \ref{table:2013_2018}.

To evaluate performance of our method on detecting poor segmentation results, we assign segmentation masks with a Jaccard score of, at least, $0.8$ as high quality and those under as low quality. The classification performance of our max consistency score is measured with AUC scores in Table \ref{table:2013_2018}. We observe that TMS-Net has large AUCs of $0.95$ for STACOM $2013$ and $0.97$ for STACOM $2018$ datasets.

\begin{table}[!ht]
\centering
\begin{minipage}{0.45 \linewidth}
\caption{Pearson Correlation Coefficients for STACOM $2013$. \label{table:Pearson2013}}
\resizebox{0.85\textwidth}{!}{%
\begin{tabular} {@{} ccccc @{} } \toprule %
&\multicolumn{2}{c}{Jaccard Indices} & \multicolumn{2}{c}{Dice Scores}\\
\cmidrule(lr){1-1} \cmidrule(lr){2-3}  \cmidrule(lr){4-4} \cmidrule(lr){4-5} 
Method & r & p & r & p\\
\cmidrule(lr){1-1} \cmidrule(lr){2-2}  \cmidrule(lr){3-3} \cmidrule(lr){4-4}  \cmidrule(lr){5-5} 
TMS-Net & \textbf{0.93} & p < .001 & \textbf{0.91} & p < .001  \\
TMS-Net3 & 0.79 & p < .001 & 0.84 &p < .001 \\
\bottomrule
\end{tabular}
}
\end{minipage}
\centering
\begin{minipage}{0.45\linewidth}
\caption{Pearson Correlation Coefficients for STACOM $2018$. \label{table:Pearson2018}}
\resizebox{0.85\textwidth}{!}{%
\begin{tabular} {@{} ccccc @{} } \toprule %
&\multicolumn{2}{c}{Jaccard Indices} & \multicolumn{2}{c}{Dice Scores}\\
\cmidrule(lr){1-1} \cmidrule(lr){2-3}  \cmidrule(lr){4-4} \cmidrule(lr){4-5} 
Method & Cor & p & Cor & p\\
\cmidrule(lr){1-1} \cmidrule(lr){2-2}  \cmidrule(lr){3-3} \cmidrule(lr){4-4}  \cmidrule(lr){5-5} 
TMS-Net & \textbf{0.96} & p < .001 & \textbf{0.95} & p < .001  \\
TMS-Net3 & 0.84 & p < .001 & 0.86 & p < .001 \\
\bottomrule
\end{tabular}
}
\end{minipage}
\end{table}

\begin{table}[!ht]
\centering
\caption{Run Time Quality Estimation Performance. Note the contribution of the specific design of TMS-Net on the performance of segmentation quality prediction at run-time, for binary classification (AUC) and Jaccard indice estimation (MAE). 
\label{table:2013_2018}}
\resizebox{0.4\textwidth}{!}{%
\begin{tabular} {@{} ccccc @{} } \toprule %
&\multicolumn{2}{c}{STACOM $2013$} & \multicolumn{2}{c}{STACOM $2018$}\\
\cmidrule(lr){1-1} \cmidrule(lr){2-3}  \cmidrule(lr){4-4} \cmidrule(lr){4-5} 
Method & AUC & MAE & AUC & MAE\\
\cmidrule(lr){1-1} \cmidrule(lr){2-2}  \cmidrule(lr){3-3} \cmidrule(lr){4-4}  \cmidrule(lr){5-5} 
TMS-Net & \textbf{0.95} & \textbf{0.13} & \textbf{0.97} & \textbf{0.11} \\
TMS-Net3 & 0.93 & 0.33 & 0.92 & 0.21 \\
\bottomrule
\end{tabular}
}
\end{table}

\section{Discussion and Conclusion} \label{sec:Discussion}

In medical image analysis, run-time performance estimation is crucial for the referral of poor quality segmentation masks to clinicians or the removal of such results from automatic image analysis pipelines. However, there are few studies on this task despite its importance \cite{roy2018inherent,robinson2019automated,hann2021ensemble}. In this study, we explored whether multi-view networks can be used for run-time quality control by viewing them as an ensemble of networks, and also proposed a max consistency score using cosine similarity to measure agreement between their decoder outputs. This study also presented a new multi-view network, TMS-Net, for robust and trustworthy medical image segmentation, which shares encoder parameters by having a single encoder versus three decoders. We observed that this specific design of TMS-Net plays an important role on its high segmentation performance, robustness to noise, aligning with previous work \cite{lee2015m,yu2020ensemble} where parameter sharing was shown to improve the performance of ensemble models. Also, we found that our design strengthens correlation between our max consistency score and run-time segmentation performance.

Our run-time quality estimation method, using our max consistency score, is easy-to-implement and unsupervised, which does not need a regression model to map features extracted from several models or MC masks to segmentation performance metrics. Although we only examined our method in detecting poor segmentation volumes as a whole, it is also possible to generate $3D$ uncertainity maps, similar to previous work by calculating variance across decoder outputs \cite{hann2021ensemble,roy2018inherent}, to more carefully investigate where the network fails. We also showed that t-SNE plots of the last encoder layer activation of TMS-Net can provide human interpretable features to understand how the network responses to non-LA and LA slices, which can be used as a tool to explain decisions of the network.

Previous work addressed the undertraining issue of segmentation models in run-time quality estimation by producing a range of quality segmentation masks with degraded versions of the same segmentation model  \cite{hann2021ensemble,robinson2019automated,valindria2017reverse}. In addition to this problem, we also examined the performance of our quality estimation method, in not-previously-explored scenarios of the presence of high anisotropy in volume images and poor imaging settings, by simulating them with Rician and engineered noise. So far, engineered noise has been mostly studied for security-related problems, which may not occur in high security of healthcare servers \cite{ma2021understanding,kaviani2022adversarial}. This study exemplifies other potential use cases of engineered noise, such as simulating data-related problems to assess model robustness to these problems. With generating poor segmentation masks with noisy images rather than by using degraded models on the original images, we were able to train a TMS-Net for segmentation once rather than training its multiple copies, in contrast to previous work \cite{hann2021ensemble,robinson2019automated,valindria2017reverse}. Therefore, we can conclude that our method has the advantage of requiring smaller memory and training time.

TMS-Net contains a trainable wavelet analysis module for image denoising to improve model robustness. Similar to our purpose, there are previous studies using wavelet and similar transformations such as shearlet transformation to improve model robustness. Ren \textit{et al.} and Abdulah \textit{et al.} respectively integrated complex shearlet scattering transform and wavelet scattering transform into a classification network to identify Covid$19$ patients from Computed Tomography (CT) images and chest X-rays \cite{ren2021detection, abdulah2022hybrid}. The goal of Ren \textit{et al.} was to obtain sparse and locally invariant image representation. Abdulah \textit{et al.} aims to improve generalisation ability of deep networks by replacing pretrained networks with classical feature extractors.

To sum up, in this work, we proposed a segmentation network, TMS-Net, and run-time segmentation quality control method for $3D$ MRI images of the left atrium, which, as long as we are aware, was not previously considered for run-time segmentation quality control. Although our latter method is developed for $3D$ images, it can be applied to $2D$ or $4D$ images by replacing TMS-Net with parameter sharing ensemble models, similar to previous work \cite{lee2015m,yu2020ensemble} and using our max consistency score on the output of the ensemble models. We believe that such a run-time quality control method has a large potential to increase trustworthiness of deep networks on medical image analysis while with the benefit of improved performance. Future work will evaluate the performance of this approach on different anatomical structures and explore different network architectures such as transformers \cite{shamshad2022transformers} in TMS-Net design.

\section*{Acknowledgments}
This work is financially supported by Bursa Technical University Scientific Research Projects Units, with the project number of $211N043$. 

\printcredits

\bibliographystyle{ieeetr} 

\bibliography{main}

\begin{thebibliography}{10}

\bibitem{cao2020application}
W.~Cao, X.~An, L.~Cong, C.~Lyu, Q.~Zhou, and R.~Guo, ``Application of deep
  learning in quantitative analysis of 2-dimensional ultrasound imaging of
  nonalcoholic fatty liver disease,'' {\em J. Med. Ultrasound}, vol.~39, no.~1,
  pp.~51--59, 2020.

\bibitem{zhou2017quantitative}
T.~Zhou, G.~Han, B.~N. Li, Z.~Lin, E.~J. Ciaccio, P.~H. Green, and J.~Qin,
  ``Quantitative analysis of patients with celiac disease by video capsule
  endoscopy: A deep learning method,'' {\em Comput. Biol. Med.}, vol.~85,
  pp.~1--6, 2017.

\bibitem{kerfoot2018left}
E.~Kerfoot, J.~Clough, I.~Oksuz, J.~Lee, A.~P. King, and J.~A. Schnabel,
  ``Left-ventricle quantification using residual u-net,'' in {\em STACOM},
  pp.~371--380, Springer, 2018.

\bibitem{uslu2021net}
F.~Uslu, M.~Varela, G.~Boniface, T.~Mahenthran, H.~Chubb, and A.~A. Bharath,
  ``La-net: A multi-task deep network for the segmentation of the left
  atrium,'' {\em IEEE TMI}, vol.~41, no.~2, pp.~456--464, 2021.

\bibitem{yang2018multiview}
G.~Yang, J.~Chen, Z.~Gao, H.~Zhang, H.~Ni, E.~Angelini, and other, ``Multiview
  sequential learning and dilated residual learning for a fully automatic
  delineation of the left atrium and pulmonary veins from late
  gadolinium-enhanced cardiac mri images,'' in {\em EMBC}, pp.~1123--1127,
  IEEE, 2018.

\bibitem{lu2021clinical}
Y.~Lu, E.~Melnick, and H.~Krumholz, ``Clinical decision support in
  cardiovascular medicine: Effectiveness, implementation barriers, and
  regulation,'' {\em MedRxiv}, 2021.

\bibitem{krittanawong2019deep}
C.~Krittanawong, K.~W. Johnson, R.~S. Rosenson, Z.~Wang, M.~Aydar, U.~Baber,
  J.~K. Min, W.~W. Tang, J.~L. Halperin, and S.~M. Narayan, ``Deep learning for
  cardiovascular medicine: a practical primer,'' {\em EHJ}, vol.~40, no.~25,
  pp.~2058--2073, 2019.

\bibitem{jia20193d}
H.~Jia, Y.~Xia, Y.~Song, D.~Zhang, H.~Huang, Y.~Zhang, and W.~Cai, ``3d
  apa-net: 3d adversarial pyramid anisotropic convolutional network for
  prostate segmentation in mr images,'' {\em IEEE TMI}, vol.~39, no.~2,
  pp.~447--457, 2019.

\bibitem{zeng2017deepem3d}
T.~Zeng, B.~Wu, and S.~Ji, ``Deepem3d: approaching human-level performance on
  3d anisotropic em image segmentation,'' {\em Bioinformatics}, vol.~33,
  no.~16, pp.~2555--2562, 2017.

\bibitem{salahuddin2022transparency}
Z.~Salahuddin, H.~C. Woodruff, A.~Chatterjee, and P.~Lambin, ``Transparency of
  deep neural networks for medical image analysis: A review of interpretability
  methods,'' {\em Comput. Biol. Med.}, vol.~140, p.~105111, 2022.

\bibitem{mehrtash2020confidence}
A.~Mehrtash, W.~M. Wells, C.~M. Tempany, P.~Abolmaesumi, and T.~Kapur,
  ``Confidence calibration and predictive uncertainty estimation for deep
  medical image segmentation,'' {\em IEEE TMI}, vol.~39, no.~12,
  pp.~3868--3878, 2020.

\bibitem{gawlikowski2021survey}
J.~Gawlikowski, C.~R.~N. Tassi, M.~Ali, J.~Lee, M.~Humt, J.~Feng, A.~Kruspe,
  R.~Triebel, P.~Jung, R.~Roscher, {\em et~al.}, ``A survey of uncertainty in
  deep neural networks,'' {\em arXiv preprint arXiv:2107.03342}, 2021.

\bibitem{hann2021deep}
E.~Hann, I.~A. Popescu, Q.~Zhang, R.~A. Gonzales, A.~Barut{\c{c}}u,
  S.~Neubauer, V.~M. Ferreira, and S.~K. Piechnik, ``Deep neural network
  ensemble for on-the-fly quality control-driven segmentation of cardiac mri t1
  mapping,'' {\em Med Image Anal}, vol.~71, p.~102029, 2021.

\bibitem{frounchi2011automating}
K.~Frounchi, L.~C. Briand, L.~Grady, Y.~Labiche, and R.~Subramanyan,
  ``Automating image segmentation verification and validation by learning test
  oracles,'' {\em Inf Softw Technol}, vol.~53, no.~12, pp.~1337--1348, 2011.

\bibitem{roy2018inherent}
A.~G. Roy, S.~Conjeti, N.~Navab, and C.~Wachinger, ``Inherent brain
  segmentation quality control from fully convnet monte carlo sampling,'' in
  {\em MICCAI}, pp.~664--672, Springer, 2018.

\bibitem{robinson2019automated}
R.~Robinson, V.~V. Valindria, W.~Bai, O.~Oktay, B.~Kainz, H.~Suzuki, M.~M.
  Sanghvi, N.~Aung, J.~M. Paiva, F.~Zemrak, {\em et~al.}, ``Automated quality
  control in image segmentation: application to the uk biobank cardiovascular
  magnetic resonance imaging study,'' {\em J Cardiovasc Magn Reson.}, vol.~21,
  no.~1, pp.~1--14, 2019.

\bibitem{hann2021ensemble}
E.~Hann, R.~A. Gonzales, I.~A. Popescu, Q.~Zhang, V.~M. Ferreira, and S.~K.
  Piechnik, ``Ensemble of deep convolutional neural networks with monte carlo
  dropout sampling for automated image segmentation quality control and robust
  deep learning using small datasets,'' in {\em MIUA}, pp.~280--293, Springer,
  2021.

\bibitem{wang2019aleatoric}
G.~Wang, W.~Li, M.~Aertsen, J.~Deprest, S.~Ourselin, and T.~Vercauteren,
  ``Aleatoric uncertainty estimation with test-time augmentation for medical
  image segmentation with convolutional neural networks,'' {\em
  Neurocomputing}, vol.~338, pp.~34--45, 2019.

\bibitem{gal2016dropout}
Y.~Gal and Z.~Ghahramani, ``Dropout as a bayesian approximation: Representing
  model uncertainty in deep learning,'' in {\em iICML}, pp.~1050--1059, PMLR,
  2016.

\bibitem{jungo2019assessing}
A.~Jungo and M.~Reyes, ``Assessing reliability and challenges of uncertainty
  estimations for medical image segmentation,'' in {\em MICCAI}, pp.~48--56,
  Springer, 2019.

\bibitem{ovadia2019can}
Y.~Ovadia, E.~Fertig, J.~Ren, Z.~Nado, D.~Sculley, S.~Nowozin, J.~Dillon,
  B.~Lakshminarayanan, and J.~Snoek, ``Can you trust your model's uncertainty?
  evaluating predictive uncertainty under dataset shift,'' {\em Adv Neural Inf
  Process Syst}, vol.~32, 2019.

\bibitem{mortazi2017cardiacnet}
A.~Mortazi, R.~Karim, K.~Rhode, J.~Burt, and U.~Bagci, ``Cardiacnet:
  segmentation of left atrium and proximal pulmonary veins from mri using
  multi-view cnn,'' in {\em MICCAI}, pp.~377--385, Springer, 2017.

\bibitem{lee2015m}
S.~Lee, S.~Purushwalkam, M.~Cogswell, D.~Crandall, and D.~Batra, ``Why m heads
  are better than one: Training a diverse ensemble of deep networks,'' {\em
  arXiv preprint arXiv:1511.06314}, 2015.

\bibitem{yu2020ensemble}
M.~Yu, V.~Cherukuri, T.~Guo, and V.~Monga, ``Ensemble dehazing networks for
  non-homogeneous haze,'' in {\em CVPR}, pp.~450--451, 2020.

\bibitem{valindria2017reverse}
V.~V. Valindria, I.~Lavdas, W.~Bai, K.~Kamnitsas, E.~O. Aboagye, A.~G. Rockall,
  D.~Rueckert, and B.~Glocker, ``Reverse classification accuracy: predicting
  segmentation performance in the absence of ground truth,'' {\em IEEE TMI},
  vol.~36, no.~8, pp.~1597--1606, 2017.

\bibitem{gheorghitua2022improving}
B.~A. Gheorghiț{\u{a}}, L.~M. Itu, P.~Sharma, C.~Suciu, J.~Wetzl, C.~Geppert,
  M.~A.~A. Ali, A.~M. Lee, S.~K. Piechnik, S.~Neubauer, {\em et~al.},
  ``Improving robustness of automatic cardiac function quantification from cine
  magnetic resonance imaging using synthetic image data,'' {\em Sci. Rep.},
  vol.~12, no.~1, pp.~1--12, 2022.

\bibitem{ronneberger2015u}
O.~Ronneberger, P.~Fischer, and T.~Brox, ``U-net: Convolutional networks for
  biomedical image segmentation,'' in {\em MICCAI}, pp.~234--241, Springer,
  2015.

\bibitem{badrinarayanan2017segnet}
V.~Badrinarayanan, A.~Kendall, and R.~Cipolla, ``Segnet: A deep convolutional
  encoder-decoder architecture for image segmentation,'' {\em IEEE PAMI},
  vol.~39, no.~12, pp.~2481--2495, 2017.

\bibitem{oktay2018attention}
O.~Oktay, J.~Schlemper, L.~L. Folgoc, M.~Lee, M.~Heinrich, K.~Misawa, K.~Mori,
  S.~McDonagh, N.~Y. Hammerla, B.~Kainz, {\em et~al.}, ``Attention u-net:
  Learning where to look for the pancreas,'' {\em arXiv preprint
  arXiv:1804.03999}, 2018.

\bibitem{zhou2018unet}
Z.~Zhou, M.~M. Rahman~Siddiquee, N.~Tajbakhsh, and J.~Liang, ``Unet++: A nested
  u-net architecture for medical image segmentation,'' in {\em DLMIA},
  pp.~3--11, Springer, 2018.

\bibitem{zhang2020dense}
Z.~Zhang, C.~Wu, S.~Coleman, and D.~Kerr, ``Dense-inception u-net for medical
  image segmentation,'' {\em Comput Methods Programs Biomed}, vol.~192,
  p.~105395, 2020.

\bibitem{siddique2021u}
N.~Siddique, S.~Paheding, C.~P. Elkin, and V.~Devabhaktuni, ``U-net and its
  variants for medical image segmentation: A review of theory and
  applications,'' {\em IEEE Access}, vol.~9, pp.~82031--82057, 2021.

\bibitem{singh20203d}
S.~P. Singh, L.~Wang, S.~Gupta, H.~Goli, P.~Padmanabhan, and B.~Guly{\'a}s,
  ``3d deep learning on medical images: a review,'' {\em Sensors}, vol.~20,
  no.~18, p.~5097, 2020.

\bibitem{cciccek20163d}
{\"O}.~{\c{C}}i{\c{c}}ek, A.~Abdulkadir, S.~S. Lienkamp, T.~Brox, and
  O.~Ronneberger, ``3d u-net: learning dense volumetric segmentation from
  sparse annotation,'' in {\em MICCAI}, pp.~424--432, Springer, 2016.

\bibitem{milletari2016v}
F.~Milletari, N.~Navab, and S.-A. Ahmadi, ``V-net: Fully convolutional neural
  networks for volumetric medical image segmentation,'' in {\em 3DV},
  pp.~565--571, IEEE, 2016.

\bibitem{kamnitsas2016deepmedic}
K.~Kamnitsas, E.~Ferrante, S.~Parisot, C.~Ledig, A.~V. Nori, A.~Criminisi,
  D.~Rueckert, and B.~Glocker, ``Deepmedic for brain tumor segmentation,'' in
  {\em BrainLes}, pp.~138--149, Springer, 2016.

\bibitem{isensee2021nnu}
F.~Isensee, P.~F. Jaeger, S.~A. Kohl, J.~Petersen, and K.~H. Maier-Hein,
  ``nnu-net: a self-configuring method for deep learning-based biomedical image
  segmentation,'' {\em Nat. Method}, vol.~18, no.~2, pp.~203--211, 2021.

\bibitem{liu2019multi}
P.~Liu, H.~Zhang, W.~Lian, and W.~Zuo, ``Multi-level wavelet convolutional
  neural networks,'' {\em IEEE Access}, vol.~7, pp.~74973--74985, 2019.

\bibitem{he2016deep}
K.~He, X.~Zhang, S.~Ren, and J.~Sun, ``Deep residual learning for image
  recognition,'' in {\em CVPR}, pp.~770--778, 2016.

\bibitem{mohan2014survey}
J.~Mohan, V.~Krishnaveni, and Y.~Guo, ``A survey on the magnetic resonance
  image denoising methods,'' {\em Biomed Signal Process Control}, vol.~9,
  pp.~56--69, 2014.

\bibitem{jifara2019medical}
W.~Jifara, F.~Jiang, S.~Rho, M.~Cheng, and S.~Liu, ``Medical image denoising
  using convolutional neural network: a residual learning approach,'' {\em J
  Supercomput}, vol.~75, no.~2, pp.~704--718, 2019.

\bibitem{li2020mri}
S.~Li, J.~Zhou, D.~Liang, and Q.~Liu, ``Mri denoising using progressively
  distribution-based neural network,'' {\em Magn. Reson. Imaging}, vol.~71,
  pp.~55--68, 2020.

\bibitem{tripathi2020cnn}
P.~C. Tripathi and S.~Bag, ``Cnn-dmri: a convolutional neural network for
  denoising of magnetic resonance images,'' {\em Pattern Recognit. Lett.},
  vol.~135, pp.~57--63, 2020.

\bibitem{mao2020multitask}
C.~Mao, A.~Gupta, V.~Nitin, B.~Ray, S.~Song, J.~Yang, and C.~Vondrick,
  ``Multitask learning strengthens adversarial robustness,'' in {\em ECCV},
  pp.~158--174, Springer, 2020.

\bibitem{goodfellow2014explaining}
I.~J. Goodfellow, J.~Shlens, and C.~Szegedy, ``Explaining and harnessing
  adversarial examples,'' {\em arXiv preprint arXiv:1412.6572}, 2014.

\bibitem{kurakin2016adversarial}
A.~Kurakin, I.~Goodfellow, S.~Bengio, {\em et~al.}, ``Adversarial examples in
  the physical world,'' 2016.

\bibitem{tobon2015benchmark}
C.~Tobon-Gomez, A.~J. Geers, J.~Peters, J.~Weese, K.~Pinto, {\em et~al.},
  ``Benchmark for algorithms segmenting the left atrium from 3d ct and mri
  datasets,'' {\em IEEE TMI}, vol.~34, no.~7, pp.~1460--1473, 2015.

\bibitem{xiong2021global}
Z.~Xiong, Q.~Xia, Z.~Hu, N.~Huang, C.~Bian, Y.~Zheng, S.~Vesal, N.~Ravikumar,
  A.~Maier, X.~Yang, {\em et~al.}, ``A global benchmark of algorithms for
  segmenting the left atrium from late gadolinium-enhanced cardiac magnetic
  resonance imaging,'' {\em Med Image Anal.}, vol.~67, p.~101832, 2021.

\bibitem{gu2019net}
Z.~Gu, J.~Cheng, H.~Fu, K.~Zhou, H.~Hao, Y.~Zhao, T.~Zhang, S.~Gao, and J.~Liu,
  ``Ce-net: Context encoder network for 2d medical image segmentation,'' {\em
  IEEE TMI}, 2019.

\bibitem{zhou2018unet++}
Z.~Zhou, M.~M. Rahman~Siddiquee, N.~Tajbakhsh, and J.~Liang, ``Unet++: A nested
  u-net architecture for medical image segmentation,'' in {\em DLMIA},
  pp.~3--11, Springer, 2018.

\bibitem{cotter2020uses}
F.~Cotter, {\em Uses of Complex Wavelets in Deep Convolutional Neural
  Networks}.
\newblock PhD thesis, University of Cambridge, 2020.

\bibitem{chen2018multi}
C.~Chen, W.~Bai, and D.~Rueckert, ``Multi-task learning for left atrial
  segmentation on ge-mri,'' in {\em STACOM}, pp.~292--301, Springer, 2018.

\bibitem{yang2018combating}
X.~Yang, N.~Wang, Y.~Wang, X.~Wang, R.~Nezafat, D.~Ni, and P.-A. Heng,
  ``Combating uncertainty with novel losses for automatic left atrium
  segmentation,'' in {\em STACOM}, pp.~246--254, Springer, 2018.

\bibitem{jia2018automatically}
S.~Jia, A.~Despinasse, Z.~Wang, H.~Delingette, X.~Pennec, P.~Ja{\"\i}s,
  H.~Cochet, and M.~Sermesant, ``Automatically segmenting the left atrium from
  cardiac images using successive 3d u-nets and a contour loss,'' in {\em
  STACOM}, pp.~221--229, Springer, 2018.

\bibitem{li2018attention}
C.~Li, Q.~Tong, X.~Liao, W.~Si, Y.~Sun, Q.~Wang, and P.-A. Heng, ``Attention
  based hierarchical aggregation network for 3d left atrial segmentation,'' in
  {\em STACOM}, pp.~255--264, Springer, 2018.

\bibitem{ma2021understanding}
X.~Ma, Y.~Niu, L.~Gu, Y.~Wang, Y.~Zhao, J.~Bailey, and F.~Lu, ``Understanding
  adversarial attacks on deep learning based medical image analysis systems,''
  {\em Pattern Recognit.}, vol.~110, p.~107332, 2021.

\bibitem{kaviani2022adversarial}
S.~Kaviani, K.~J. Han, and I.~Sohn, ``Adversarial attacks and defenses on ai in
  medical imaging informatics: A survey,'' {\em Expert Syst. Appl}, p.~116815,
  2022.

\bibitem{ren2021detection}
Q.~Ren, B.~Zhou, L.~Tian, and W.~Guo, ``Detection of covid-19 with ct images
  using hybrid complex shearlet scattering networks,'' {\em IEEE J Biomed
  Health Inform}, vol.~26, no.~1, pp.~194--205, 2021.

\bibitem{abdulah2022hybrid}
H.~Abdulah, B.~Huber, H.~Abdallah, L.~L. Palese, H.~Soltanian-Zadeh, and D.~L.
  Gatti, ``A hybrid pipeline for covid-19 screening incorporating lungs
  segmentation and wavelet based preprocessing of chest x-rays,'' {\em
  medRxiv}, 2022.

\bibitem{shamshad2022transformers}
F.~Shamshad, S.~Khan, S.~W. Zamir, M.~H. Khan, M.~Hayat, F.~S. Khan, and H.~Fu,
  ``Transformers in medical imaging: A survey,'' {\em arXiv preprint
  arXiv:2201.09873}, 2022.

\bibitem{mallat1989theory}
S.~G. Mallat, ``A theory for multiresolution signal decomposition: the wavelet
  representation,'' {\em IEEE PAMI}, vol.~11, no.~7, pp.~674--693, 1989.

\end{thebibliography}

\appendix

\section{Appendix}
\subsection{Wavelet Analysis For Denoising} \label{sec:WaveletAnalysis}
TMS-Net uses wavelet analysis in its design so we will summarise its main points. A typical wavelet based image denoising method includes the subsequent steps of wavelet decomposition, wavelet coefficients analysis and wavelet reconstruction \cite{mohan2014survey}. Wavelet decomposition, $W$, and reconstruction, $W^{-1}$, can be performed with $F_{l},F_{h}=W(F;f) $ and $F=W^{-1}(F_{l},F_{h};f) $ consecutively, where $f$ represents non-trainable kernels, such as Haar filters, used for the wavelet analysis of a feature set of $F\in R^{H \times W}$. $F_{l} \in R^{1 \times \frac{H}{2} \times \frac{W}{2}}$ and $F_{h}  \in R^{ 3 \times \frac{H}{2} \times \frac{W}{2}}$ respectively represent low frequency features, $F_{l}=F_{ll}$, and the set of high frequency features, $F_{h}=\left \{F_{lh}, F_{hl}, F_{hh}\right\}$. $W$ and $H$ correspond to feature width and height consecutively. 

$2D$ discrete Haar filters are frequently used for their better edge detection property \cite{mohan2014survey}; $f_{ll}=\frac{1}{2}\begin{bmatrix}
1 & 1 \\
1 & 1
\end{bmatrix}$, $f_{lh}=\frac{1}{2}\begin{bmatrix}
-1 & -1 \\
1 & 1
\end{bmatrix}$, $f_{hl}=\frac{1}{2}\begin{bmatrix}
-1 & 1 \\
-1 & 1
\end{bmatrix}$, $f_{hh}=\frac{1}{2}\begin{bmatrix} \nonumber
1 & -1 \\
-1 & 1
\end{bmatrix}$ \cite{mallat1989theory}. $f_{ll}$ acts as average pooling and $f_{lh}$, $f_{hl}$ and $f_{hh}$ respectively generates edge-like features along rows, columns and diagonals. The wavelet decomposition naturally downsamples an input image by a factor of $2$ and the wavelet reconstruction does the opposite effect. Liu \textit{et al} used them as pooling and unpooling operations in deep networks, by calling them \say{wavelet pooling, $W_{p}$} and \say{wavelet unpooling, $W_{up}$} \cite{liu2019multi}.  We will also follow the same terminology in our work.

\subsection{Effect of Three View Engineered, and Rician Noise to Segmentation Masks} \label{sec:3viewAdvers}

\begin{figure*}[!ht]
\centering
\includegraphics[width=1\textwidth]{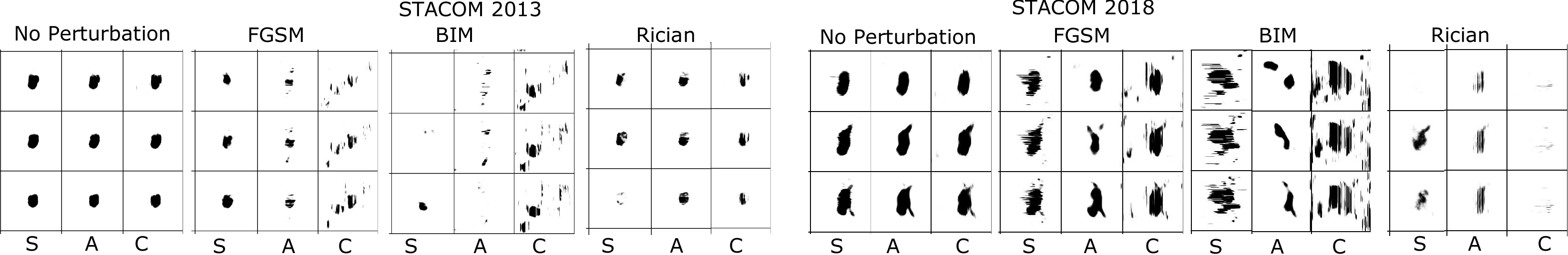}
\caption{Example outputs of TMS-Net decoders $S$, $A$ and $C$ before and after engineered noises of FGSM and BIM methods and Rician noise. The outputs of decoders are converted to the axial view for visual comparison. $\epsilon=0.04$ for FGSM and BIM engineered noise and $\sigma =15$ for Rician noise.} \label{fig:ExampleOutputs}
\end{figure*}

We examine decoder outputs of TMS-Net in the case of three view corruptions with FGSM and BIM methods, and Rician noise in Figure \ref{fig:ExampleOutputs}. As understood from the figure, large similarities across decoder outputs present for the original images, decrease after large engineered and Rician noise to a great extent.

\bio{}
\endbio

\end{document}